\documentclass[apj]{emulateapj}
\usepackage{apjfonts}
\usepackage{lscape}

\shorttitle{Stellar multiplicity in Cha~I}
\shortauthors{Lafreni\`{e}re et al.}

\newcommand{\msun}{\ensuremath{M_{\odot}}}

\begin{document}

\title{A Multiplicity Census of Young Stars in Chamaeleon~I}

\author{David Lafreni\`ere\altaffilmark{1},
Ray Jayawardhana\altaffilmark{1},
Alexis Brandeker\altaffilmark{2},
Mirza Ahmic\altaffilmark{1},
Marten H. van Kerkwijk\altaffilmark{1}
}
\altaffiltext{1}{Department of Astronomy and Astrophysics, University of Toronto, 50 St. George Street, Toronto, ON, M5S 3H4, Canada}
\altaffiltext{2}{Stockholm Observatory, AlbaNova University Center, SE-105 91 Stockholm, Sweden}
\email{lafreniere@astro.utoronto.ca}

\begin{abstract} 
We present the results of a multiplicity survey of 126 stars spanning $\sim$0.1--3~\msun\ in the $\sim$2-Myr-old Chamaeleon~I star-forming region, based on adaptive optics imaging with the ESO Very Large Telescope. Our observations have revealed 30 binaries and 6 triples, of which 19 and 4, respectively, are new discoveries. The overall multiplicity fraction we find for Cha~I ($\sim$30\%) is similar to those reported for other dispersed young associations, but significantly higher than seen in denser clusters and the field, for comparable samples. Both the frequency and the maximum separation of Cha~I binaries decline with decreasing mass, while the mass ratios approach unity; conversely, tighter pairs are more likely to be equal mass. We confirm that brown dwarf companions to stars are rare, even at young ages at wide separations. Based on follow-up spectroscopy of two low-mass substellar companion candidates, we conclude that both are likely background stars. The overall multiplicity fraction in Cha~I is in rough agreement with numerical simulations of cloud collapse and fragmentation, but its observed mass dependence is less steep than predicted. The paucity of higher-order multiples, in particular, provides a stringent constraint on the simulations, and seems to indicate a low level of turbulence in the prestellar cores in Cha~I.
\end{abstract} 

\section{Introduction}

The ubiquity of multiple stellar systems, both in the field \citep[e.g.][]{duquennoy91} and in star-forming regions \citep[e.g.][]{ghez97}, clearly indicates that multiplicity is a principal channel of star formation. Accordingly, the multiplicity properties of a stellar population -- such as the fraction of binary or higher-order multiple systems, the orbital separation and mass ratio distributions of companions, and the dependence of these on stellar mass -- should be good tracers of the specific mechanism by which the stars formed, and possibly also of the characteristics of their birth environment \citep[e.g.][]{goodwin04, bate03}. Beyond this primordial imprint, many of these properties can be further modified by dynamical interactions or decay of the newborn systems over a few million years \citep[e.g.][]{sterzik98, sterzik03, kroupa95, kroupa98}, a process that may also be environment dependent and that could introduce spatial variations of multiplicity properties within a given cluster. Studies of stellar multiplicity can thus provide valuable insight into the global process of star formation and early evolution, especially if multiplicity properties can be compared across a wide range of stellar mass, for different ages, and for different environments.

The multiplicity properties of stellar systems in the Solar neighborhood have been studied extensively \citep[e.g.][]{duquennoy91, fischer92,
delfosse04}, at least for $\sim$0.1--2~\msun\ stars. The picture is not yet as complete for substellar objects, but progress is being made rapidly \citep[see][and references therein]{burgasser07}. For pre-main-sequence stars, since they usually lie more than 100~pc away from Earth, observational advances have been closely tied to the development of large aperture telescopes and high angular resolution imaging techniques. The first speckle imaging surveys of a few star-forming regions, carried out in the early 1990s, revealed binary frequencies that appeared to be significantly higher than in the field \citep{ghez93, leinert93}. This early realization prompted observations of more young
clusters, and the ensuing results suggested that multiplicity may vary with the environment \citep{ghez97, padgett97, petr98, kohler98, simon99, duchene99, kohler00}. While dispersed Taurus-like associations show an overabundance of binaries by a factor $\sim$2 in comparison to the field, the binary fraction in dense Trapezium-like clusters appears to be in good agreement with the field value \citep{petr98, simon99}. Whether this dichotomy is a direct outcome of the star formation process or the result of stronger dynamical evolution in denser environments remains unclear \citep{petr98, duchene99, mathieu00, brandeker06}.

Typically, owing to the brightness requirements inherent to speckle imaging, many of the early surveys neglected the fainter, very low mass (VLM) members of young associations and clusters. However, a number of recent multiplicity studies have specifically targeted brown dwarfs and VLM stars in star-forming regions \citep[e.g.][]{kraus05,kraus06,bouy06,ahmic07,konopacky07}. The young VLM binaries found in these surveys share several characteristics with their counterparts in the field: lower binary fraction than for higher mass stars, lower average orbital separation, and a preference for mass ratios near unity \citep{close03, burgasser03, delfosse04, burgasser07}. But, the current observations also hint at a higher incidence of wide ($>$30~AU) VLM binaries in star-forming regions in comparison with the field \citep[e.g.][]{konopacky07, jayawardhana06}.

The compendium of high angular resolution imaging data of pre-main-sequence stars now covers a wide range of environments, ages and masses. However, the various surveys often have different resolution and contrast limits, and their samples are drawn from different mass regimes. As a result, determining the mass dependence of multiplicity and comparing the properties of different star-forming environments are no easy tasks. This is rather unfortunate, especially given that recent theoretical efforts, which have focused on modeling the collapse and fragmentation of prestellar cores and the early dynamical interactions of small-N stellar clusters \citep[e.g.][]{sterzik03, delgado04, goodwin04}, have produced results that could be compared directly with observable quantities. Thus, we have embarked upon an ambitious program to obtain a comprehensive and detailed picture of the multiplicity properties of pre-mains-sequence stars across a wide mass range, based
on a dataset of homogeneous quality and sensitivity for several star-forming regions.

Here we present the results of an adaptive optics (AO) multiplicity survey of 126 stars spanning $\sim$0.1--3~\msun\ in the nearby (160 pc; \citealp{whittet97}), young (2 Myr; \citealp{luhman04} and references therein) Chamaeleon I (hereafter Cha I) cloud. \citet{ahmic07} presented an analysis of the subsample of VLM targets from this survey while \citet{damjanov07} investigated circumstellar disks of Cha~I members, including the effect of companions on disks. The present paper reports the results for the entire sample observed, with a particular emphasis on the multiplicity properties as a function of stellar mass.

The structure of the paper is as follows. The target sample and the observations are described in \S\ref{sect:obs}, and the data analysis performed is detailed in \S\ref{sect:analysis}. The multiple systems identified and their measured properties are presented in \S\ref{sect:results}. The multiplicity characteristics of our sample, including the multiplicity fraction and its dependence on mass, the mass ratio of companions as a function of primary mass and orbital separation, the binary separations as a function of mass, and the stability of the triple systems, are discussed in \S\ref{sect:discussion} in relation to other stellar populations and to numerical simulations of star formation. Finally, concluding remarks follow in \S\ref{sect:conclusion}.

\section{Observations and data reduction}\label{sect:obs}

The target sample was built upon the recent compilation of Cha~I members made by \citet{luhman04}. The membership of these stars was established based on space velocity, H-R diagram position, infrared excess emission, lithium abundance, and stellar activity indicators.
The only selection criterion applied to the list of \citet{luhman04} was a magnitude cut at $K$=13, the faint magnitude limit of the IR wave sensor of the AO system used for the observations; this removed only 8 out of the 158 known Cha~I members and does not introduce a significant bias. We have observed 126 out of the remaining 150 Cha~I members, our sample is thus a good representation of the entire population of Cha~I. The basic properties of the stars observed are indicated in Table~\ref{tbl:obs}.\footnote{Six other stars were observed during the observing run but are not considered in this paper as the AO correction achieved was extremely poor (Ced~110~IRS~6, Cha~H$\alpha$~10, Cha~H$\alpha$~11, IRN, ISO~217, and ISO~220); these stars are omitted from Table~\ref{tbl:obs}.} The $K$-band magnitudes, spectral types and effective temperatures of the targets were taken from \citet{luhman04}.

The observations were obtained with the Nasmyth Adaptive Optics System (NAOS) and the CONICA camera at the 8.2m telescope Yepun, part of the European Southern Observatory's Very Large Telescope (program 076.C-0579). The data were acquired in visitor mode over three consecutive nights starting on March 24, 2006, with the exception of the source Cha~H$\alpha$~2, which was observed in service mode on 2005 March 25 as part of a preliminary run (program 075.C-0042). The high-resolution lens of CONICA was used for all observations, yielding a pixel scale of 13.27~mas and a field of view of 13.6\arcsec\ $\times$ 13.6\arcsec. For the majority of targets, the N90C10 dichroic was used to direct 90\% of the light to the AO wave front sensor and 10\% to the science camera, but whenever possible the VIS dichroic was used to take advantage of its higher throughput toward CONICA at near-IR wavelengths. All targets have been observed in the $K$ band using either the $K_{\rm s}$ or the narrow-band 2.17~$\mu$m filter, and most targets have been further observed in the $H$ band using either the $H$ or the narrow-band 1.64~$\mu$m filter. A typical imaging sequence consisted of six exposures divided over two dither positions separated by $\sim$5.5\arcsec. Observing parameters for each target are indicated in Table~\ref{tbl:obs}. The data were reduced in a standard manner using the NAOS-CONICA pipeline. As an indication of image quality, the $K$-band point-spread function (PSF) full-width-at-half-maximum (FWHM) is indicated in Table~\ref{tbl:obs}. The image quality was highly variable over the three nights of observations, and the median $K$-band FWHM observed is 0.086\arcsec. For reference, the $K$-band diffraction-limited FWHM is 0.056\arcsec.

The images of the stars CHXR~18N and T54 showed faint candidate companions that, if truly bound, would have masses well below the substellar limit. However, being faint and located at relatively large separations from their primary (2.7\arcsec and 2.4\arcsec, resp.), the probabilities that these sources are actually unrelated background stars are significant. Based on the analysis presented in \S\ref{sect:prob}, these probabilities are 0.33 and 0.40, respectively. To assess whether or not these candidates are truly bound companions and to clarify their nature, we have obtained follow-up imaging and spectroscopy at the VLT on 2007 May 13 for T54 [278.C-5070(A)] and on 2007 June 18 for CHXR~18N [279.C-5017(A)]. The imaging follow-up was obtained in $J$, $H$, and $K_{\rm s}$ using the same instrumental configuration as our main survey and data reduction was carried out as described above. For T54, the total exposure times were 7.5~min, 9~min, and 4.8~min in $J$, $H$, and $K_{\rm s}$, respectively. For CHXR~18N, they were 9~min, 2.8~min, and 2~min, respectively. The spectroscopic observations were acquired using the VIS wavefront sensor, a 40\arcsec$\times$0.172\arcsec\ slit, and the S27-4-SH spectroscopic mode of NACO, covering the wavelength range 1.37--1.84~$\mu$m. Exposures at two different positions along the slit were obtained two times in an ABBA pattern, with each position observed 63$\times$8~s for T54 and 9$\times$18~s for CHXR~18N, totaling 4$\times$63$\times$8~s = 34~min integration on T54 and 4$\times$9$\times$18~s = 11~min on CHXR~18N. For the binary T54 and its candidate companion as well as for CHXR~18N and its candidate companion, the slit was aligned with the primary and candidate companion; for T54, this means that the tight secondary component was out of the slit. Immediately after each observation, a calibration star was observed with the same settings and at a similar airmass to correct for atmospheric and instrumental transmission. The B4 star HIP~53024 was used for T54 and the solar-like G3 star HIP~88299 was used for CHXR~18N.

\onecolumngrid
\LongTables
\begin{deluxetable}{lcccccccccccccc}
\tablewidth{0pt}
\tabletypesize{\scriptsize}
\tablecolumns{15}
\tablecaption{Target properties and observation log \label{tbl:obs}}
\tablehead{
\colhead{} & \colhead{RA} & \colhead{DEC} & \colhead{$K_{\rm s}$} & \colhead{Spectral} & \colhead{$T_{\rm eff}$\tablenotemark{a}} & \colhead{Mass\tablenotemark{b}} & \multicolumn{2}{c}{$t_{\rm exp}$ (s)} & \colhead{Dich.\tablenotemark{c}} & \colhead{FWHM\tablenotemark{d}} & \multicolumn{4}{c}{Detection limits ($K_{\rm s}$)} \\
\cline{8-9}\cline{12-15}
\colhead{Name} & \colhead{(J2000.0)} & \colhead{(J2000.0)} & \colhead{(mag)} & \colhead{type\tablenotemark{a}} & \colhead{(K)} & \colhead{($M_\odot$)} & \colhead{$K_{\rm s}$} & \colhead{$H$} & \colhead{} & \colhead{(\arcsec)} & \colhead{0.1\arcsec} & \colhead{0.25\arcsec} & \colhead{0.5\arcsec} & \colhead{1.0\arcsec}
}
\startdata
CHSM 17173 & 11 10 22.27 & -76 25 13.8 & 12.45 & M8 & 2710 & 0.03 & 240 & $-$ & 1 & 0.131  & 13.7 & 15.2 & 15.8 & 15.8 \\
Cha H$\alpha$ 1 & 11 07 16.69 & -77 35 53.3 & 12.17 & M7.75 & 2752 & 0.06 & 120 & 120 & 1 & 0.144  & 13.2 & 14.2 & 14.5 & 14.5 \\
Cha H$\alpha$ 7 & 11 07 37.76 & -77 35 30.8 & 12.42 & M7.75 & 2752 & 0.06 & 240 & $-$ & 1 & 0.133  & 13.5 & 14.9 & 15.5 & 15.6 \\
Cha H$\alpha$ 12 & 11 06 38.00 & -77 43 09.1 & 11.81 & M6.5 & 2935 & 0.12 & 120 & 120 & 1 & 0.119  & 13.1 & 14.5 & 15.0 & 15.0 \\
ISO 138 & 11 08 18.50 & -77 30 40.8 & 13.04 & M6.5 & 2935 & 0.12 & 240 & $-$ & 1 & 0.121  & 14.2 & 15.0 & 15.2 & 15.1 \\
CHSM 1982 & 11 04 10.60 & -76 12 49.0 & 12.12 & M6 & 2990 & 0.14 & 120 & 120 & 1 & 0.092  & 13.9 & 15.0 & 15.2 & 15.2 \\
ISO 252 & 11 10 41.41 & -77 20 48.0 & 12.27 & M6 & 2990 & 0.14 & 120 & $-$ & 1 & 0.135  & 13.4 & 14.7 & 15.1 & 15.1 \\
CHSM 10862 & 11 07 46.56 & -76 15 17.5 & 12.33 & M5.75 & 3024 & 0.14 & 120 & $-$ & 1 & 0.123  & 13.6 & 14.6 & 15.0 & 15.0 \\
Cha H$\alpha$ 6 & 11 08 39.52 & -77 34 16.7 & 11.04 & M5.75 & 3024 & 0.14 & 120 & 120 & 1 & 0.105  & 12.9 & 14.9 & 15.5 & 15.5 \\
Hn 13 & 11 10 55.97 & -76 45 32.6 &  9.91 & M5.75 & 3024 & 0.14 & 120 & 108 & 1 & 0.072  & 12.3 & 14.3 & 15.3 & 15.9 \\
ISO 147 & 11 08 26.51 & -77 15 55.1 & 12.35 & M5.75 & 3024 & 0.14 & 240 & $-$ & 1 & 0.144  & 13.4 & 14.6 & 15.2 & 15.3 \\
CHXR 84 & 11 12 03.27 & -76 37 03.4 & 10.78 & M5.5 & 3058 & 0.15 & 120 & 120 & 1 & 0.091  & 13.0 & 15.2 & 16.0 & 16.1 \\
Cha H$\alpha$ 13 & 11 08 17.03 & -77 44 11.8 & 10.67 & M5.5 & 3058 & 0.15 & 120 & 120 & 1 & 0.096  & 12.9 & 15.3 & 16.2 & 16.3 \\
Cha H$\alpha$ 3 & 11 07 52.26 & -77 36 57.0 & 11.10 & M5.5 & 3058 & 0.15 & 120 & 120 & 1 & 0.096  & 13.3 & 15.7 & 16.4 & 16.4 \\
Cha H$\alpha$ 4 & 11 08 18.96 & -77 39 17.0 & 11.02 & M5.5 & 3058 & 0.15 & 120 & 120 & 1 & 0.086  & 13.3 & 15.5 & 16.1 & 16.1 \\
Cha H$\alpha$ 5 & 11 08 24.11 & -77 41 47.4 & 10.71 & M5.5 & 3058 & 0.15 & 120 & 120 & 1 & 0.113  & 12.5 & 14.7 & 15.6 & 15.5 \\
Cha H$\alpha$ 9 & 11 07 18.61 & -77 32 51.7 & 11.80 & M5.5 & 3058 & 0.15 & 120 & 120 & 1 & 0.099  & 13.4 & 14.9 & 15.4 & 15.4 \\
Hn 12W & 11 10 28.52 & -77 16 59.6 & 10.78 & M5.5 & 3058 & 0.15 & 120 & 120 & 1 & 0.095  & 12.7 & 14.7 & 15.4 & 15.5 \\
ISO 165 & 11 08 54.97 & -76 32 41.1 & 11.44 & M5.5 & 3058 & 0.15 & 120 & 120 & 1 & 0.132  & 12.5 & 13.7 & 14.4 & 14.5 \\
ISO 235 & 11 10 07.85 & -77 27 48.1 & 11.34 & M5.5 & 3058 & 0.15 & 120 & 120 & 1 & 0.103  & 13.0 & 14.8 & 15.7 & 15.8 \\
ISO 28 & 11 03 41.87 & -77 26 52.0 & 11.69 & M5.5 & 3058 & 0.15 & 120 & 120 & 1 & 0.098  & 13.3 & 14.8 & 15.3 & 15.3 \\
CHXR 15 & 11 05 43.00 & -77 26 51.8 & 10.23 & M5.25 & 3091 & 0.16 & 30 & 30 & 1 & 0.086  & 12.6 & 14.2 & 14.5 & 14.5 \\
CHXR 78C & 11 08 54.22 & -77 32 11.6 & 11.22 & M5.25 & 3091 & 0.16 & 30 & 30 & 1 & 0.112  & 12.9 & 13.6 & 13.5 & 13.5 \\
Cha H$\alpha$ 2\tablenotemark{e} & 11 07 42.45 & -77 33 59.4 & 10.68 & M5.25 & 3091 & 0.16 & 1080 & $-$ & 1 & 0.087  & 12.4 & 13.9 & 14.5 & 14.9 \\
C7-1 & 11 09 42.60 & -77 25 57.9 & 10.55 & M5 & 3125 & 0.17 & 30 & 30 & 1 & 0.126  & 12.0 & 13.5 & 13.7 & 13.6 \\
Hn 2 & 11 03 47.64 & -77 19 56.3 &  9.99 & M5 & 3125 & 0.17 & 30 & 30 & 1 & 0.072  & 12.7 & 14.9 & 15.5 & 15.5 \\
ISO 143 & 11 08 22.38 & -77 30 27.7 & 11.10 & M5 & 3125 & 0.17 & 30 & 30 & 1 & 0.098  & 12.9 & 14.1 & 14.3 & 14.3 \\
T50 & 11 12 09.85 & -76 34 36.6 &  9.84 & M5 & 3125 & 0.17 & 30 & 30 & 1 & 0.069  & 12.9 & 15.3 & 15.9 & 15.9 \\
Hn 7 & 11 09 05.13 & -77 09 58.1 & 10.96 & M4.75 & 3161 & 0.17 & 30 & 30 & 1 & 0.090  & 13.0 & 14.4 & 14.6 & 14.5 \\
ISO 250 & 11 10 36.45 & -77 22 13.2 & 10.67 & M4.75 & 3161 & 0.17 & 30 & 30 & 1 & 0.100  & 12.6 & 13.9 & 14.1 & 14.1 \\
B35 & 11 07 21.43 & -77 22 11.8 & 10.93 & -- & -- & 0.18\tablenotemark{f} & 30 & $-$ & 1 & 0.113  & 12.5 & 13.6 & 13.8 & 13.8 \\
Hn 5 & 11 06 41.81 & -76 35 49.0 & 10.12 & M4.5 & 3198 & 0.18 & 30 & 30 & 1 & 0.081  & 12.4 & 14.3 & 14.9 & 15.0 \\
ISO 274 & 11 11 22.61 & -77 05 53.9 & 10.69 & M4.5 & 3198 & 0.18 & 30 & 30 & 1 & 0.092  & 12.7 & 14.4 & 14.8 & 14.8 \\
T12 & 11 02 55.05 & -77 21 50.8 & 10.45 & M4.5 & 3198 & 0.18 & 30 & 30 & 1 & 0.089  & 12.8 & 14.8 & 15.2 & 15.1 \\
T55 & 11 13 33.57 & -76 35 37.4 & 10.73 & M4.5 & 3198 & 0.18 & 30 & 30 & 1 & 0.098  & 13.0 & 14.7 & 14.8 & 14.7 \\
CHXR 60 & 11 13 29.71 & -76 29 01.2 & 10.58 & M4.25 & 3234 & 0.19 & 30 & 30 & 1 & 0.091  & 12.4 & 13.9 & 14.2 & 14.1 \\
CHXR 74 & 11 06 57.33 & -77 42 10.7 & 10.21 & M4.25 & 3234 & 0.19 & 30 & 30 & 1 & 0.095  & 11.7 & 13.0 & 13.6 & 13.7 \\
CHXR 76 & 11 07 35.19 & -77 34 49.3 & 10.95 & M4.25 & 3234 & 0.19 & 30 & 30 & 1 & 0.094  & 13.0 & 14.5 & 14.8 & 14.7 \\
2M1110-7722\tablenotemark{h} & 11 10 34.81 & -77 22 05.3 & 10.03 & M4 & 3270 & 0.20 & 30 & 30 & 1 & 0.083  & 12.2 & 14.3 & 15.1 & 15.1 \\
Hn 21W & 11 14 24.54 & -77 33 06.2 & 10.65 & M4 & 3270 & 0.20 & 120 & 120 & 1 & 0.091  & 12.5 & 14.7 & 15.8 & 16.0 \\
ISO 52 & 11 04 42.58 & -77 41 57.1 & 10.64 & M4 & 3270 & 0.20 & 30 & 30 & 1 & 0.097  & 12.6 & 14.1 & 14.5 & 14.4 \\
CHXR 62 & 11 14 15.65 & -76 27 36.4 & 10.12 & M3.75 & 3306 & 0.22 & 30 & 30 & 1 & 0.087  & 11.9 & 13.2 & 13.6 & 13.7 \\
T10 & 11 00 40.22 & -76 19 28.1 & 10.87 & M3.75 & 3306 & 0.22 & 30 & 30 & 1 & 0.094  & 13.1 & 14.3 & 14.4 & 14.5 \\
T34 & 11 08 16.49 & -77 44 37.2 & 10.02 & M3.75 & 3306 & 0.22 & 30 & 30 & 1 & 0.072  & 12.7 & 14.6 & 15.0 & 15.1 \\
CHXR 12 & 11 03 56.83 & -77 21 33.0 &  9.71 & M3.5 & 3342 & 0.23 & 30 & 30 & 1 & 0.082  & 12.1 & 14.1 & 14.8 & 14.9 \\
CHXR 22E & 11 07 13.30 & -77 43 49.9 &  9.99 & M3.5 & 3342 & 0.23 & 30 & 30 & 1 & 0.098  & 11.7 & 13.9 & 14.7 & 14.6 \\
CHXR 26\tablenotemark{e} & 11 07 36.87 & -77 33 33.5 &  9.35 & M3.5 & 3342 & 0.23 & 30 & 30 & 1 & 0.085  & 11.5 & 14.0 & 14.8 & 14.7 \\
Hn 18 & 11 13 24.46 & -76 29 22.7 & 10.80 & M3.5 & 3342 & 0.23 & 30 & 30 & 1 & 0.081  & 13.4 & 15.2 & 15.5 & 15.4 \\
B43 & 11 09 47.42 & -77 26 29.1 & 10.24 & M3.25 & 3379 & 0.24 & 30 & 30 & 1 & 0.090  & 12.2 & 14.3 & 15.1 & 15.1 \\
CHXR 73\tablenotemark{k} & 11 06 28.77 & -77 37 33.2 & 10.70 & M3.25 & 3379 & 0.24 & 30 & 30 & 1 & 0.099  & 12.6 & 14.1 & 14.5 & 14.5 \\
Hn 10E & 11 09 46.21 & -76 34 46.4 & 10.05 & M3.25 & 3379 & 0.24 & 30 & 30 & 1 & 0.076  & 12.6 & 14.5 & 15.0 & 15.1 \\
Hn 4 & 11 05 14.67 & -77 11 29.1 &  9.61 & M3.25 & 3379 & 0.24 & 30 & 30 & 1 & 0.081  & 12.1 & 13.9 & 14.3 & 14.5 \\
T5 & 10 57 42.20 & -76 59 35.7 &  9.25 & M3.25 & 3379 & 0.24 & 30 & 30 & 1 & 0.072  & 12.2 & 13.7 & 14.3 & 14.9 \\
CHXR 21 & 11 07 11.49 & -77 46 39.4 &  9.66 & M3 & 3415 & 0.26 & 30 & 30 & 1 & 0.098  & 11.6 & 13.8 & 14.6 & 14.7 \\
CHXR 71 & 11 02 32.65 & -77 29 13.0 & 10.13 & M3 & 3415 & 0.26 & 30 & 30 & 1 & 0.070  & 12.9 & 14.8 & 15.1 & 15.1 \\
T16 & 11 04 57.01 & -77 15 56.9 & 10.41 & M3 & 3415 & 0.26 & 30 & 30 & 1 & 0.086  & 12.7 & 14.6 & 15.2 & 15.1 \\
T22 & 11 06 43.47 & -77 26 34.4 &  9.39 & M3 & 3415 & 0.26 & 30 & 30 & 1 & 0.076  & 12.0 & 14.5 & 15.4 & 15.4 \\
B53 & 11 14 50.32 & -77 33 39.0 &  9.55 & M2.75 & 3451 & 0.27 & 30 & 30 & 1 & 0.079  & 11.9 & 14.0 & 14.7 & 14.8 \\
CHXR 57 & 11 13 20.13 & -77 01 04.5 & 10.01 & M2.75 & 3451 & 0.27 & 30 & 30 & 1 & 0.069  & 13.2 & 15.5 & 15.9 & 15.9 \\
CHXR 59 & 11 13 27.37 & -76 34 16.6 &  9.63 & M2.75 & 3451 & 0.27 & 30 & 30 & 1 & 0.078  & 12.1 & 13.7 & 14.3 & 14.5 \\
Cam2-19 & 11 06 15.45 & -77 37 50.1 & 10.25 & M2.75 & 3451 & 0.27 & 30 & 30 & 1 & 0.094  & 12.3 & 14.2 & 14.8 & 14.6 \\
CHXR 48 & 11 11 34.75 & -76 36 21.1 &  9.80 & M2.5 & 3488 & 0.29 & 30 & 30 & 1 & 0.078  & 12.4 & 14.6 & 15.2 & 15.2 \\
T25 & 11 07 19.15 & -76 03 04.8 &  9.77 & M2.5 & 3488 & 0.29 & 30 & 30 & 1 & 0.075  & 12.5 & 14.9 & 15.5 & 15.5 \\
T30 & 11 07 58.09 & -77 42 41.3 &  9.89 & M2.5 & 3488 & 0.29 & 30 & 30 & 1 & 0.079  & 12.4 & 14.6 & 15.2 & 15.2 \\
CHXR 9C & 11 01 18.75 & -76 27 02.5 &  8.99 & M2.25 & 3524 & 0.32 & 30 & 30 & 1 & 0.082  & 11.4 & 13.3 & 13.5 & 13.5 \\
T39\tablenotemark{g} & 11 09 11.72 & -77 29 12.5 &  8.96 & M2 & 3560 & 0.35 & 48 & 30 & 1 & 0.079  & 11.4 & 13.9 & 14.6 & 14.5 \\
T43 & 11 09 54.08 & -76 29 25.3 &  9.25 & M2 & 3560 & 0.35 & 30 & 30 & 1 & 0.072  & 12.2 & 14.7 & 15.4 & 15.4 \\
T47 & 11 10 49.60 & -77 17 51.8 &  9.18 & M2 & 3560 & 0.35 & 30 & 30 & 1 & 0.072  & 11.6 & 13.9 & 14.9 & 14.9 \\
T49 & 11 11 39.66 & -76 20 15.2 &  8.87 & M2 & 3560 & 0.35 & 48 & 30 & 1 & 0.090  & 10.9 & 13.1 & 13.9 & 14.0 \\
CHXR 14S & 11 04 52.85 & -76 25 51.5 &  9.75 & M1.75 & 3596 & 0.37 & 30 & 30 & 1 & 0.078  & 12.1 & 14.4 & 15.1 & 15.0 \\
T20 & 11 05 52.61 & -76 18 25.6 &  9.34 & M1.5 & 3632 & 0.40 & 30 & 30 & 1 & 0.069  & 12.3 & 14.8 & 15.7 & 15.7 \\
T23 & 11 06 59.07 & -77 18 53.6 & 10.00 & M1.5 & 3632 & 0.40 & 30 & 30 & 1 & 0.070  & 12.9 & 15.1 & 15.6 & 15.5 \\
C1-6 & 11 09 22.67 & -76 34 32.0 &  8.67 & M1.25 & 3669 & 0.42 & 30 & 30 & 1 & 0.083  & 10.7 & 13.3 & 14.5 & 14.6 \\
CHXR 30B & 11 07 57.31 & -77 17 26.2 &  9.95 & M1.25 & 3669 & 0.42 & 30 & 30 & 1 & 0.076  & 12.3 & 14.6 & 15.5 & 15.5 \\
CHXR 40 & 11 09 40.07 & -76 28 39.2 &  8.96 & M1.25 & 3669 & 0.42 & 48 & 30 & 1 & 0.090  & 10.9 & 12.7 & 13.4 & 13.6 \\
CHXR 79 & 11 09 18.13 & -76 30 29.2 &  9.06 & M1.25 & 3669 & 0.42 & 30 & 30 & 1 & 0.086  & 11.3 & 13.6 & 14.6 & 14.7 \\
ISO 126 & 11 08 02.98 & -77 38 42.6 &  8.30 & M1.25 & 3669 & 0.42 & 48 & 30 & 1 & 0.079  & 10.6 & 12.8 & 13.9 & 14.4 \\
CHXR 54 & 11 12 42.10 & -76 58 40.0 &  9.50 & M1 & 3705 & 0.44 & 30 & 30 & 1 & 0.075  & 12.2 & 14.7 & 15.4 & 15.4 \\
T27\tablenotemark{e} & 11 07 28.26 & -76 52 11.9 &  9.52 & M1 & 3705 & 0.44 & 30 & 30 & 1 & 0.079  & 11.9 & 13.8 & 14.4 & 14.6 \\
T28 & 11 07 43.66 & -77 39 41.1 &  8.26 & M1 & 3705 & 0.44 & 48 & 30 & 1 & 0.072  & 10.7 & 13.1 & 14.3 & 14.8 \\
T48 & 11 10 53.34 & -76 34 32.0 & 10.04 & M1 & 3705 & 0.44 & 30 & 30 & 1 & 0.078  & 12.6 & 14.5 & 15.1 & 15.1 \\
T53 & 11 12 30.93 & -76 44 24.1 &  9.12 & M1 & 3705 & 0.44 & 30 & 30 & 1 & 0.069  & 12.0 & 14.6 & 15.7 & 15.8 \\
T24 & 11 07 12.07 & -76 32 23.2 &  9.38 & M0.5 & 3778 & 0.48 & 30 & 30 & 1 & 0.076  & 12.0 & 14.6 & 15.4 & 15.5 \\
T3\tablenotemark{e} & 10 55 59.73 & -77 24 39.9 &  8.69 & M0.5 & 3778 & 0.48 & 30 & 30 & 1 & 0.106  & 10.2 & 12.5 & 14.1 & 14.6 \\
T38 & 11 08 54.64 & -77 02 13.0 &  9.46 & M0.5 & 3778 & 0.48 & 30 & $-$ & 1 & 0.171  & 10.3 & 11.4 & 12.0 & 12.1 \\
T4 & 10 56 30.45 & -77 11 39.3 &  8.63 & M0.5 & 3778 & 0.48 & 30 & $-$ & 1 & 0.075  & 11.0 & 13.7 & 15.0 & 15.2 \\
T56 & 11 17 37.01 & -77 04 38.1 &  9.23 & M0.5 & 3778 & 0.48 & 30 & 30 & 1 & 0.069  & 12.2 & 14.6 & 15.4 & 15.4 \\
C1-25 & 11 09 41.93 & -76 34 58.4 & 10.00 & -- & -- & 0.49\tablenotemark{f} & 30 & 30 & 1 & 0.136  & 11.2 & 12.6 & 13.1 & 13.1 \\
CHXR 33 & 11 08 40.69 & -76 36 07.8 &  9.28 & M0 & 3850 & 0.51 & 30 & 30 & 1 & 0.069  & 12.1 & 14.6 & 15.7 & 15.7 \\
CHXR 49NE & 11 11 54.00 & -76 19 31.1 &  9.23 & M0 & 3850 & 0.51 & 30 & 30 & 1 & 0.069  & 12.3 & 14.2 & 14.7 & 14.9 \\
T45A & 11 10 04.69 & -76 35 45.3 &  9.24 & M0 & 3850 & 0.51 & 30 & 30 & 1 & 0.069  & 11.9 & 14.5 & 15.7 & 15.9 \\
CHXR 14N & 11 04 51.00 & -76 25 24.1 &  9.60 & K8 & 3955 & 0.55 & 30 & 30 & 1 & 0.079  & 12.0 & 14.4 & 15.2 & 15.2 \\
CHXR 30A & 11 08 00.03 & -77 17 30.5 &  9.09 & K8 & 3955 & 0.55 & 30 & 30 & 1 & 0.085  & 11.3 & 13.5 & 14.3 & 14.4 \\
CHXR 68A & 11 18 20.24 & -76 21 57.6 &  8.87 & K8 & 3955 & 0.55 & 48 & 30 & 1 & 0.076  & 11.4 & 13.4 & 14.1 & 14.3 \\
Hn 11 & 11 10 03.69 & -76 33 29.2 &  9.44 & K8 & 3955 & 0.55 & 30 & 30 & 1 & 0.085  & 11.7 & 14.1 & 15.1 & 15.1 \\
T31\tablenotemark{g} & 11 08 01.49 & -77 42 28.9 &  6.96 & K8 & 3955 & 0.55 & 30 & 30 & 1 & 0.075  &  9.8 & 12.0 & 13.4 & 14.6 \\
T35 & 11 08 39.05 & -77 16 04.2 &  9.11 & K8 & 3955 & 0.55 & 30 & 30 & 1 & 0.069  & 11.8 & 14.1 & 15.1 & 15.3 \\
T45\tablenotemark{e} & 11 09 58.74 & -77 37 08.9 &  7.97 & K8 & 3955 & 0.55 & 30 & 30 & 1 & 0.089  &  9.9 & 12.0 & 13.3 & 14.0 \\
T46 & 11 10 07.04 & -76 29 37.7 &  8.45 & K8 & 3955 & 0.55 & 48 & 30 & 1 & 0.069  & 11.2 & 13.6 & 15.0 & 15.7 \\
T7 & 10 59 01.09 & -77 22 40.7 &  8.62 & K8 & 3955 & 0.55 & 30 & 30 & 1 & 0.072  & 11.2 & 13.9 & 15.1 & 15.3 \\
CHXR 37 & 11 09 17.70 & -76 27 57.8 &  8.70 & K7 & 4060 & 0.59 & 30 & 30 & 1 & 0.070  & 10.8 & 13.9 & 15.3 & 15.6 \\
Cam2-42 & 11 09 37.78 & -77 10 41.1 &  9.16 & K7 & 4060 & 0.59 & 30 & 30 & 1 & 0.090  & 11.3 & 13.7 & 14.8 & 14.9 \\
C1-2 & 11 09 55.06 & -76 32 41.0 &  9.67 & -- & -- & 0.64\tablenotemark{f} & 30 & 30 & 1 & 0.108  & 11.1 & 12.9 & 13.8 & 13.8 \\
CHXR 18N & 11 11 46.32 & -76 20 09.2 &  7.77 & K6 & 4205 & 0.65 & 30 & $-$ & 1 & 0.072  & 10.3 & 12.7 & 14.3 & 15.3 \\
CHXR 20 & 11 06 45.10 & -77 27 02.3 &  8.88 & K6 & 4205 & 0.65 & 30 & 30 & 1 & 0.095  & 10.7 & 12.7 & 13.6 & 13.7 \\
CHXR 28\tablenotemark{e} & 11 07 55.89 & -77 27 25.8 &  7.69 & K6 & 4205 & 0.65 & 30 & 30 & 1 & 0.092  &  9.2 & 11.0 & 12.5 & 13.6 \\
T11 & 11 02 24.91 & -77 33 35.7 &  8.20 & K6 & 4205 & 0.65 & 48 & 30 & 1 & 0.082  & 10.7 & 13.6 & 15.2 & 15.7 \\
T40 & 11 09 23.79 & -76 23 20.8 &  8.24 & K6 & 4205 & 0.65 & 48 & 30 & 1 & 0.069  & 11.0 & 13.7 & 15.3 & 16.0 \\
ISO 237 & 11 10 11.42 & -76 35 29.3 &  8.62 & K5.5 & 4278 & 0.68 & 30 & 30 & 1 & 0.079  & 11.0 & 13.6 & 14.9 & 15.4 \\
T14 & 11 04 09.09 & -76 27 19.4 &  8.66 & K5 & 4350 & 0.71 & 30 & 30 & 1 & 0.069  & 11.6 & 14.3 & 15.5 & 15.8 \\
T42 & 11 09 53.41 & -76 34 25.5 &  6.46 & K5 & 4350 & 0.71 & 30 & 30 & 1 & 0.107  &  7.9 &  9.8 & 11.2 & 11.9 \\
T44 & 11 10 00.11 & -76 34 57.9 &  6.08 & K5 & 4350 & 0.71 & 48 & $-$ & 1 & 0.121  &  7.2 &  9.2 & 10.7 & 11.5 \\
CHXR 55 & 11 12 43.00 & -76 37 04.9 &  9.29 & K4.5 & 4470 & 0.77 & 30 & 30 & 1 & 0.069  & 12.3 & 15.1 & 16.0 & 16.0 \\
T51\tablenotemark{e} & 11 12 24.41 & -76 37 06.4 &  8.00 & K3.5 & 4660 & 0.88 & 48 & 30 & 1 & 0.092  &  9.7 & 11.6 & 13.3 & 14.5 \\
CHXR 47\tablenotemark{e} & 11 10 38.02 & -77 32 39.9 &  8.28 & K3 & 4730 & 0.96 & 48 & 30 & 1 & 0.091  & 10.5 & 12.4 & 13.5 & 14.3 \\
T8 & 10 59 06.99 & -77 01 40.4 &  7.31 & K2 & 4900 & 1.30 & 31 & $-$ & 2 & 0.078  &  9.8 & 12.0 & 13.4 & 14.9 \\
T6\tablenotemark{e} & 10 58 16.77 & -77 17 17.1 &  7.76 & K0 & 5250 & 1.96 & 30 & 30 & 1 & 0.072  & 10.0 & 12.5 & 14.3 & 15.3 \\
T52 & 11 12 27.72 & -76 44 22.3 &  6.84 & G9 & 5410 & 2.23 & 31 & $-$ & 2 & 0.110  &  8.2 &  9.9 & 11.4 & 13.0 \\
T54\tablenotemark{e} & 11 12 42.69 & -77 22 23.1 &  7.88 & G8 & 5520 & 2.41 & 30 & 30\tablenotemark{j} & 2 & 0.083  & 10.2 & 12.4 & 13.8 & 15.3 \\
T33A+B\tablenotemark{e} & 11 08 15.10 & -77 33 53.2 &  6.18 & G7 & 5630 & 2.51 & 30 & 30 & 1 & 0.086  &  8.4 & 10.4 & 11.8 & 12.8 \\
T21 & 11 06 15.41 & -77 21 56.8 &  6.42 & G5 & 5770 & 2.54 & 30 & 30 & 1 & 0.108  &  7.8 &  9.7 & 11.2 & 12.1 \\
T26\tablenotemark{e} & 11 07 20.74 & -77 38 07.3 &  6.22 & G2 & 5860 & 2.57 & 90\tablenotemark{i} & 45\tablenotemark{j} & 2 & 0.072  &  8.8 & 10.9 & 12.6 & 14.3 \\
HD 93828 & 10 46 37.95 & -77 36 03.6 &  7.48 & F0 & 7200 & 2.87 & 31 & 31 & 2 & 0.082  &  9.9 & 12.1 & 13.7 & 15.5 \\
T41\tablenotemark{e} & 11 09 50.03 & -76 36 47.7 &  7.15 & B9 & 10500 & $>3$  & 31 & 31 & 2 & 0.094  &  8.9 & 11.0 & 12.5 & 13.6 \\
HD 96675 & 11 05 57.81 & -76 07 48.9 &  6.97 & B6.5 & 13500 & $>3$  & 31 & 31 & 2 & 0.099  &  8.5 & 10.4 & 11.6 & 12.6 \\
T32 & 11 08 03.30 & -77 39 17.4 &  5.94 & B9.5 & 10010 & $>3$  & 90\tablenotemark{i} & 31 & 2 & 0.072  &  8.4 & 10.7 & 12.5 & 14.0 \\
\enddata
\tablenotetext{a}{From \citet{luhman04}.}
\tablenotetext{b}{Unless stated otherwise, estimate based on $T_{\rm eff}$ and the models of \citet{dantona97}; see text for more detail.}
\tablenotetext{c}{Dichroic used: (1) N90C10 or (2) VIS.}
\tablenotetext{d}{In $K$-band.}
\tablenotetext{e}{Object was known as a binary from previous studies.}
\tablenotetext{f}{Estimate based on absolute $K$-band magnitude and the models of \citet{dantona97}; see \S\ref{sect:masses} for more detail.}
\tablenotetext{g}{Object was known as a triple from previous studies.}
\tablenotetext{h}{Complete name is 2MASS J11103481-7722053.}
\tablenotetext{i}{This observation was done in the NB$_{2.17}$ filter.}
\tablenotetext{j}{This observation was done in the NB$_{1.64}$ filter.}
\tablenotetext{k}{We did not detect the substellar companion to this source reported by \citet{luhman06} as it is below our detection limits.}
\end{deluxetable}
\twocolumngrid

The spectroscopic data were reduced in a standard way, except that an unfortunate ghost at the location of the companion position in the data of CHXR~18N data prevented us from using the B position as sky for A. Instead, we fit a linear function to the sky along the slit. This procedure did not leave any visible residuals from the sky OH lines. The spectra were extracted using an optimal extraction algorithm, fitting a parametric PSF allowed to vary with wavelength. Comparisons with spectra extracted by simply summing up all flux in a large aperture yielded consistent results. Wavelength calibration was made using night sky OH emission lines and the atlas of \citet{rousselot00}. The spectra were corrected for extinction, using the $R=3.1$ parametric formula from \citet{fitzpatrick99} and assuming the same extinction as measured for the primaries by \citet[][$A_J$=0.60 for T54 and $A_J$=0.14 for CHXR\,18N]{luhman04}.

The motivation for choosing the 172~mas wide slit was the need to ensure that the companion candidates would be located in the slit using blind acquisitions, given their relatively large contrasts and separations with respect to their primaries. At a separation of $\sim$2.5\arcsec, a small uncertainty of $\sim$1\degr\ in position angle would translate into a $\sim$50~mas uncertainty in position, comparable to the PSF FWHM. In addition, as the AO correction varies strongly with wavelength, the resulting PSF is wavelength-dependent; this effect was amplified by the relatively high airmass (1.7) of our observations, which follows from the southern declination of the targets (-77\degr). The wide slit used thus considerably simplifies the wavelength-dependent relative flux calibration, but also adds the complication that the spectral resolution becomes PSF dependent, in our case $R$ = 200--500, and with a complicated line-spread function. This limits the accuracy to which we can remove telluric absorption lines by dividing by the calibration stars (and multiplying by a model), as the spectral resolution for the science and calibration observations differ. To mitigate this effect, we convolved the calibration spectra with a Gaussian until the telluric residuals were minimized (although not entirely removed). Since our main interest lies in the shape of the spectrum, however, and not in individual absorption lines, we did not try to optimize further the telluric feature removal.

\section{Analysis}\label{sect:analysis}

\subsection{Identification and measurements of multiple systems}

Identification of multiple systems was first done by visual inspection of the images. Then, for well-separated multiple systems ($>0.5$\arcsec), subtraction of the stellar signal using the PSF of another component was performed to search for additional tight, low-mass companions to each of the components. While this approach led to very small residuals, no additional companion was found. For apparently single stars, PSF subtraction using the images of other targets was attempted but yielded little improvement due to the large variability of the PSF throughout the three nights and again, no additional companion was found. Although the PSFs of a few targets are elongated, which could indicate unresolved binarity, it is difficult to make concrete assertions as there are several instances of intrinsically elongated PSFs in our data (i.e. elongation visible in both components of a binary system).

The angular separation, position angle, and magnitude difference between the components of multiple systems were determined using a custom PSF extraction/subtraction {\em IDL} routine. For well-separated systems, this routine determines the parameters that minimize the residual noise after subtraction of one component from another. For close-separation systems, the routine first reconstructs the PSF, using the then best guess of the separation and flux ratio of the components, from the parts of each component's PSF that are uncontaminated by the other component, and then determines the parameters that minimize the residual noise after subtraction of the reconstructed PSF from the image. This process is iterated until convergence is reached. The random errors on each parameter were estimated from the dispersion of the measurements made on the six individual images of each sequence. By comparing the FWHM and encircled energy of the primary stars and their companions as a function of angular separation, we did not find evidence for significant anisoplanatism effects below 6\arcsec, and thus we did not apply any correction to the measured photometry for such effects. 

\subsection{Mass estimates}\label{sect:masses}

The masses of all targets were estimated from their effective temperatures and the predictions of pre-main sequence evolution models; this approach was preferred over a determination based on luminosity as it is less affected by unresolved binarity or uncertainties in extinction or excess emission. Choosing which pre-main-sequence evolution model to use is a delicate task as these models are highly uncertain at young ages and there are significant differences between the different ones available. For the present study, as one of its goals is to investigate the variations of the properties of multiple systems as a function of mass, the masses should ideally be estimated in a consistent manner across the entire range of masses observed ($\sim$0.02--4~\msun\ for primaries and companions). We have thus opted for the models of \citet{dantona97}\footnote{We have used the 1998 update of these models for masses in the range 0.017--0.9~\msun.}, whose wide mass coverage (0.017--3~\msun) enables determination of the masses of all but three of the primary stars observed and all  but one of the companions found (see \S\ref{sect:results}). For completeness, a discussion of the effects of using other evolution models to determine masses is presented in \S\ref{sect:models}. For three targets for which the spectral type (hence effective temperature) is unknown, the masses were determined based on absolute $K$-band magnitude. To do so, the luminosities from the models of \citet{dantona97} were converted to $K$ magnitudes using the bolometric corrections and colors tabulated in \citet{kenyon95} for $T_{\rm eff}\ge 3000$, and in \citet{bessell91} for $T_{\rm eff} < 3000$. The estimated masses of all targets are indicated in Table~\ref{tbl:obs}. An age of 2~Myr \citep{luhman04} was assumed for all targets to produce these estimates. This simple assumption introduces additional uncertainties as in reality there is likely an age spread among the Cha~I members ($\sim$1--5~Myr, see Fig.~15 of \citealp{luhman04}). By comparing the masses obtained for ages of 1 or 5~Myr to those obtained for an age of 2~Myr using the models of \citet{dantona97}, we estimate that these uncertainties are of the order of 10--15\%.

The mass ratios of the secondary (and tertiary) components of multiple systems were estimated based on their $K_{\rm s}$ contrast with respect to their primary. More precisely, the observed $K$-band {\em contrast} of the companion was added to the {\em model} $K$-band magnitude of the primary (i.e. based on its effective temperature), and the corresponding companion mass was then obtained from the models; this approach is less sensitive to uncertainties in distance and extinction than one relying on observed absolute magnitudes. However, the observed flux ratios may be slightly different from the true photospheric flux ratios if there is excess emission from circumstellar material around one or both components of a system. For reference, we point out that an error of 0.2~mag on the photospheric flux ratio would typically lead to a difference of roughly 15\% in mass ratio. As discussed before for the primary mass, the uncertainties on the ages of the systems also introduce uncertainties on the estimated mass ratios. We estimate that these uncertainties are of the order of 15--20\%, using the evolution models of \citet{dantona97}. For the tertiary component of CHXR~68 and the secondary component of Hn~21W, we have used the effective temperatures (3524~K and 3024~K, resp.) reported in \citet{luhman04} to derive their masses.

\subsection{Detection limits}

The detection limits as a function of angular separation from the target stars were obtained by calculating the residual noise in annuli centered on the target after subtraction of an azimuthally symmetric PSF profile and convolution of the residual image by a circular aperture of radius 0.04\arcsec. The detection limits were determined as 5 times the residual noise. For multiple systems, the secondary and/or tertiary component was subtracted from the image using an analytic PSF model prior to the calculation of the detection limits. Since at the smallest separations the variance of the residual noise may be underestimated by this procedure as there are too few independent spatial elements within the corresponding annuli, we have required that the detection limits within a radius of 2 FWHM be no better than the PSF intensity profile. As stated before, we found no evidence for significant anisoplanatism within the angular separation range probed, and thus we did not apply any correction to the detection limits for such an effect.

The $K$-band detection limits are indicated in Table~\ref{tbl:obs} for four angular separations. We provide detection limits at 0.1\arcsec\ even for targets for which the PSF FWHM exceeds 0.1\arcsec\ since for these targets the PSF is very smooth and very finely sampled by the 0.013\arcsec\ pixels such that companions with separations slightly below one FWHM could be detected, provided they are above the primary PSF intensity profile. The median $K$-band detection limits are 12.2~mag at 0.1\arcsec, 14.1~mag at 0.25\arcsec, 14.8~mag at 0.5\arcsec, and 15.0~mag at 1\arcsec\ and beyond. In terms of contrast, the median limits are 2.2, 4.4, 5.1, and 5.3~mag, respectively. Mass ratio limits have been calculated for each target based on their $K_{\rm s}$ contrast limit in the manner described above. For each of four mass bins and two angular separations, the minimum mass ratios that could have been detected for 50\% ($q_{50\%}$) and 90\% ($q_{90\%}$) of the targets are reported in Table~\ref{tbl:qlim}; the latter value can be used as an estimate of our completeness limit. The higher mass ratio limits at small separations for the last mass bin of Table~\ref{tbl:qlim} arise from a combination of higher speckle noise for these brighter stars and a brightening of $\sim$2--3~\msun\ stars at $\sim$2~Myr as they settle on the zero-age main sequence and the CNO cycle reaches equilibrium (L. Siess 2007, private communication). 

\begin{deluxetable}{lcccc}
\tablewidth{0pt}
\tablecolumns{5}
\tablecaption{Mass ratio limits \label{tbl:qlim}}
\tablehead{
\colhead{Mass bin} & \multicolumn{4}{c}{Mass ratio limit ($q_{50\%}/q_{90\%}$)} \\
\cline{2-5}
\colhead{($M_\odot$)} & \colhead{0.1\arcsec} & \colhead{0.25\arcsec} & \colhead{0.5\arcsec} & \colhead{1\arcsec}
}
\startdata
0.10--0.23 & 0.23/0.44 & 0.12/0.24 & 0.09/0.21 & 0.09/0.21 \\
0.23--0.55 & 0.13/0.20 & 0.04/0.08 & 0.03/0.05 & 0.03/0.06 \\
0.55--1.28 & 0.11/0.23 & 0.04/0.06 & 0.02/0.04 & 0.02/0.03 \\
1.28--3.00 & 0.29/0.60 & 0.04/0.14 & 0.02/0.03 & 0.01/0.02
\enddata
\end{deluxetable}

\section{Results}\label{sect:results}

\subsection{Resolved multiple systems}

\begin{figure}[t!]
\epsscale{1}
\plotone{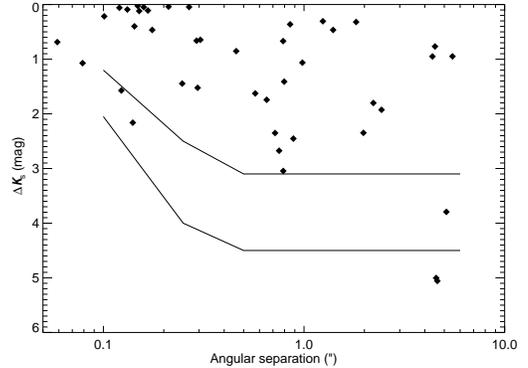}
\caption{\label{fig:comp} Separation and contrast, relative to their primary star, of all the companions identified in this study. The two curves indicate detection limits that were obtained for at least 90\% ({\it top}) and 50\% ({\it bottom}) of the targets.}
\end{figure}

\begin{figure}[b!]
\epsscale{1}
\plotone{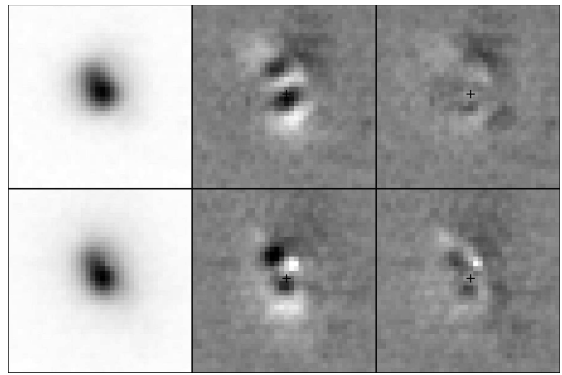}
\caption{\label{fig:2m1110} $K_{\rm s}$- ({\it top}) and $H$-band ({\it bottom}) AO images of the tight binary system 2M1110-7722 ({\it first column}). The other columns show the same images after subtraction of the best-fit single-component ({\it second column}) and two-component ({\it third column}) analytic PSF models consisting of Gaussian$+$Moffat functions; the display intensity scale is $\pm10\%$ of the primary PSF peak. The PSF models are allowed to be asymmetric. Each panel is 0.5\arcsec\ on a side. The smaller residuals after subtraction of the two-component PSF model clearly indicate that this is a binary system rather than a single star with an elongated PSF.}
\end{figure}

\begin{figure*}
\epsscale{0.8}
\plotone{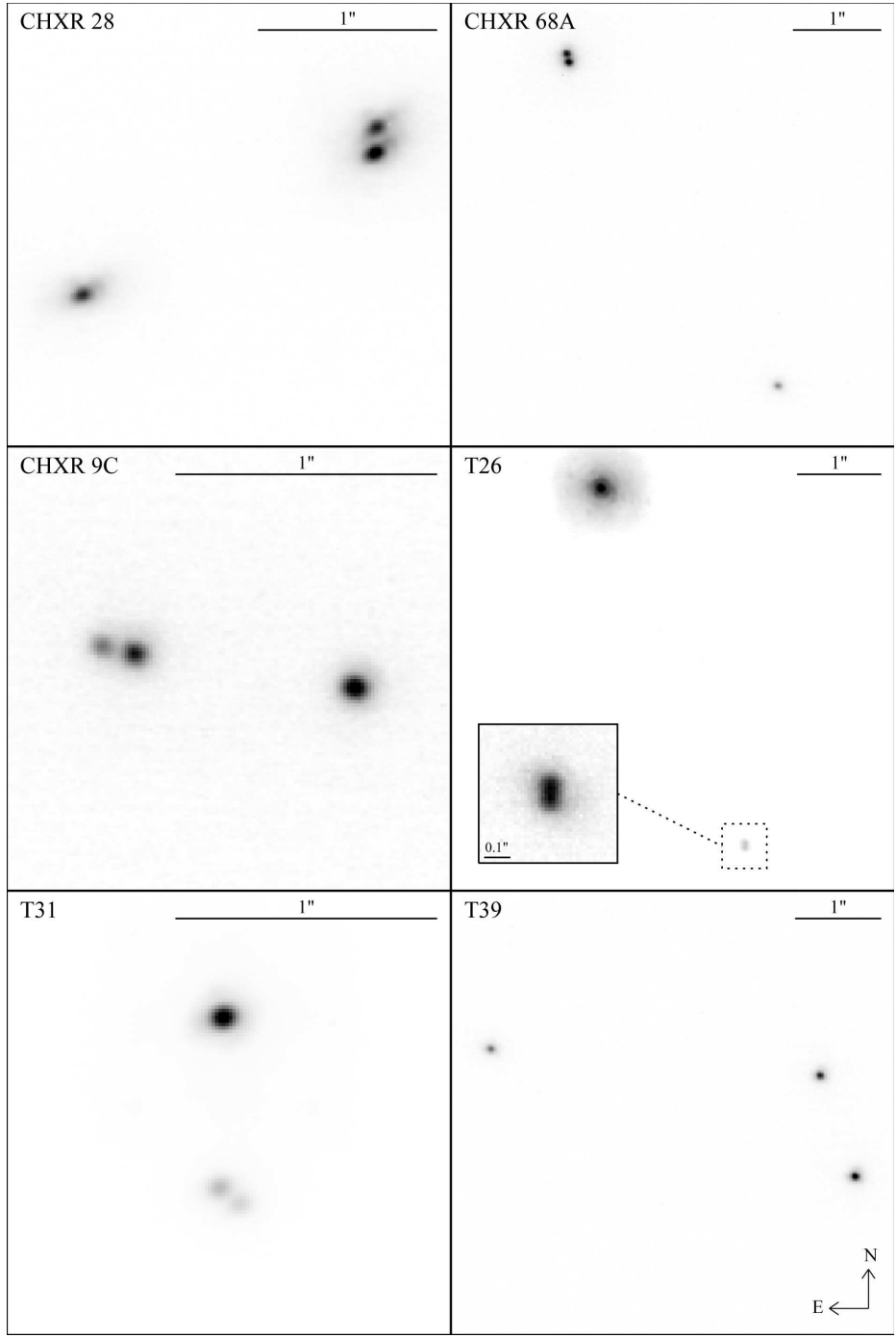}
\caption{\label{fig:triples} Triples systems identified in our survey. The intensity scale is linear, except for T26 which is shown on a logarithmic intensity scale.}
\end{figure*}

We have identified a total of 30 candidate binary systems and 7 candidate triple systems among the 126 targets of our sample. No higher-order multiple system was found. The majority of the binaries and all but two (T31 and T39) of the triples are new discoveries. The triple systems CHXR~28, CHXR~68A and T26 were previously known binaries for which one of the components was itself resolved into a tight binary. The properties of the binary and triple systems observed are summarized in Tables~\ref{tbl:mult} and \ref{tbl:tri}. Only the random measurement errors are reported in Table~\ref{tbl:mult}, a systematic uncertainty of 0.226\% in separation and 0.5\degr\ in position angle should be added \citep{masciadri03}. The separation and $K_{\rm s}$ contrast of all companions identified are shown in Figure~\ref{fig:comp}. Figure~\ref{fig:2m1110} presents images of the tightest binary found in our survey, 2M1110-7722, which is marginally resolved with a separation of 0.059\arcsec\ and contrast of 0.69~mag in $K_{\rm s}$. Images of the triples systems identified are shown in Figure~\ref{fig:triples}. The faint companion candidate around the binary T54 and the faint companion candidate around CHXR~18N are omitted from table~\ref{tbl:tri} and figures~\ref{fig:comp},\,\ref{fig:triples} as follow-up observations have indicated that they are background stars (see \S\ref{sect:followup}). They are however listed in table~\ref{tbl:mult} for completeness.

The triple systems were named based on hierarchical levels. The components of the wide pairs were assigned the capital letters {\em A,B} and the components of the tight pairs were assigned the lower case letters {\em a,b}; the brighter component of each pair bears the letter {\em A} or {\em a}. The configuration and properties of the triple systems, split into tight and wide pairs, are indicated in Table~\ref{tbl:tri}. For the wide pair, the tight subsystem is considered as a single entity whose mass is equal to the sum of the masses of its components, and the separation of the wide system is calculated with respect to the center of mass of the tight subsystem. Among our sample of triple systems, the two

\onecolumngrid
\begin{deluxetable}{lccccccc}
\tablewidth{0pt}
\tabletypesize{\footnotesize}
\tablecolumns{7}
\tablecaption{Measurements and properties of multiple systems \label{tbl:mult}}
\tablehead{
\colhead{Name} & \colhead{Separation\tablenotemark{a}} & \colhead{Position Angle\tablenotemark{b}} & \colhead{$\Delta$$K_{\rm s}$} & \colhead{$\Delta$$H$} & \colhead{$q$\tablenotemark{c}} & \colhead{$P^\prime$} \\
\colhead{} & \colhead{(\arcsec)} & \colhead{(\degr)} & \colhead{(mag)} & \colhead{(mag)} & \colhead{} & \colhead{(yr)}
}
\startdata
\multicolumn{7}{c}{\dotfill Binaries \dotfill} \\
2M1110-7722 & $ 0.059\pm0.002$ & $ 30.1\pm1.5$ & $0.69\pm0.04$ & $0.57\pm0.19$ & 0.46 &    50 \\
B53 & $ 0.295\pm0.001$ & $235.2\pm0.1$ & $1.52\pm0.03$ & $1.55\pm0.01$ & 0.24 &   600 \\
CHXR 15 & $ 0.304\pm0.001$ & $199.9\pm0.2$ & $0.65\pm0.01$ & $0.63\pm0.01$ & 0.53 &   700 \\
CHXR 26 & $ 1.396\pm0.001$ & $ 33.8\pm0.1$ & $0.47\pm0.01$ & $0.45\pm0.01$ & 0.51 &  6000 \\
CHXR 30A & $ 0.459\pm0.001$ & $ 23.2\pm0.1$ & $0.85\pm0.01$ & $0.95\pm0.02$ & 0.41 &   700 \\
CHXR 37 & $ 0.079\pm0.001$ & $ 81.0\pm1.1$ & $1.07\pm0.01$ & $1.09\pm0.04$ & 0.32 &    50 \\
CHXR 40 & $ 0.151\pm0.002$ & $ 66.1\pm0.6$ & $0.12\pm0.03$ & $0.19\pm0.03$ & 0.89 &   130 \\
CHXR 47 & $ 0.175\pm0.001$ & $334.8\pm0.2$ & $0.47\pm0.05$ & $0.46\pm0.01$ & 0.73 &   120 \\
CHXR 49NE & $ 0.267\pm0.001$ & $ 53.2\pm0.1$ & $0.05\pm0.01$ & $0.04\pm0.01$ & 0.96 &   300 \\
CHXR 59 & $ 0.148\pm0.001$ & $347.4\pm0.2$ & $0.02\pm0.01$ & $0.06\pm0.03$ & 0.97 &   160 \\
CHXR 62 & $ 0.120\pm0.002$ & $135.1\pm1.0$ & $0.06\pm0.09$ & $0.12\pm0.09$ & 0.93 &   130 \\
CHXR 71 & $ 0.572\pm0.001$ & $ 70.0\pm0.1$ & $1.63\pm0.01$ & $1.75\pm0.01$ & 0.23 &  1600 \\
CHXR 79 & $ 0.885\pm0.001$ & $211.2\pm0.1$ & $2.45\pm0.02$ & $2.10\pm0.03$ & 0.11 &  2000 \\
Cha H$\alpha$ 2 & $ 0.167\pm0.001$ & $ 40.4\pm0.2$ & $0.11\pm0.02$ & \nodata & 0.77 &   300 \\
Hn 13 & $ 0.132\pm0.001$ & $318.2\pm0.2$ & $0.09\pm0.01$ & $0.17\pm0.01$ & 0.81 &   190 \\
Hn 21W & $ 5.495\pm0.004$ & $ 69.2\pm0.1$ & $0.95\pm0.02$ & $0.91\pm0.02$ & 0.71 & 40000 \\
Hn 4 & $ 0.211\pm0.001$ & $296.1\pm0.3$ & $0.04\pm0.01$ & $0.02\pm0.18$ & 0.95 &   300 \\
ISO 126 & $ 0.292\pm0.001$ & $232.4\pm0.2$ & $0.66\pm0.01$ & $0.49\pm0.01$ & 0.46 &   400 \\
T21 & $ 0.140\pm0.006$ & $126.1\pm2.3$ & $2.16\pm0.12$ & $2.05\pm0.11$ & 0.33 &    60 \\
T27 & $ 0.787\pm0.001$ & $ 13.5\pm0.1$ & $0.67\pm0.01$ & $0.50\pm0.01$ & 0.46 &  1800 \\
T3 & $ 2.216\pm0.002$ & $289.0\pm0.1$ & $1.80\pm0.01$ & $1.17\pm0.02$ & 0.16 &  9000 \\
T33A+B & $ 2.434\pm0.001$ & $284.8\pm0.1$ & $1.93\pm0.04$ & $0.66\pm0.12$ & 0.38 &  4000 \\
T41 & $ 0.788\pm0.001$ & $326.7\pm0.1$ & $3.05\pm0.03$ & $3.33\pm0.06$ & \nodata  & \nodata  \\
T43 & $ 0.796\pm0.001$ & $352.2\pm0.1$ & $1.41\pm0.01$ & $1.43\pm0.01$ & 0.22 &  2000 \\
T45 & $ 0.752\pm0.006$ & $ 52.2\pm0.4$ & $2.67\pm0.02$ & $2.34\pm0.03$ & 0.09 &  1700 \\
T46 & $ 0.123\pm0.001$ & $241.6\pm0.6$ & $1.57\pm0.02$ & $1.38\pm0.08$ & 0.18 &   110 \\
T5 & $ 0.159\pm0.001$ & $342.9\pm0.1$ & $0.05\pm0.01$ & $0.04\pm0.01$ & 0.95 &   190 \\
T51 & $ 1.977\pm0.001$ & $162.5\pm0.1$ & $2.35\pm0.01$ & $2.34\pm0.03$ & 0.10 &  6000 \\
T54\tablenotemark{d} & $ 0.247\pm0.001$ & $246.5\pm0.2$ & $1.45\pm0.04$ & $1.54\pm0.01$ & 0.52 &   130 \\
T6 & $ 5.122\pm0.003$ & $122.9\pm0.1$ & $3.79\pm0.03$ & $3.90\pm0.06$ & 0.04 & 16000 \\
\\[-2ex]
\multicolumn{7}{c}{\dotfill Triples \dotfill} \\
CHXR~28~Aa,Ab & $ 0.143\pm0.001$ & $356.6\pm1.0$ & $0.40\pm0.05$ & $0.40\pm0.04$ & 0.73 & 100 \\
CHXR~28~Aa,B  & $ 1.818\pm0.003$ & $115.9\pm0.1$ & $0.32\pm0.04$ & $0.31\pm0.10$ & 0.77 & \nodata \\[0.8ex]

CHXR 68A~Aa,Ab & $ 0.101\pm0.001$ & $ 13.6\pm1.5$ & $0.22\pm0.02$ & $0.33\pm0.07$ & 0.84 & 60 \\
CHXR~68A~Aa,B  & $ 4.367\pm0.005$ & $213.0\pm0.1$ & $0.95\pm0.12$ & $0.27\pm0.57$ & 0.58 & \nodata \\[0.8ex]

CHXR~9C~A,Ba & $ 0.852\pm0.001$ & $ 81.3\pm0.1$ & $0.36\pm0.04$ & $0.45\pm0.11$ & 0.63 & \nodata \\
CHXR 9C~Ba,Bb& $ 0.130\pm0.002$ & $ 76.1\pm0.4$ & $0.70\pm0.04$ & $0.67\pm0.11$ & 0.46 & 170 \\[0.8ex]

T26~A,Ba & $ 4.557\pm0.007$ & $202.1\pm0.1$ & $5.00\pm0.03$ & $4.55\pm0.12$ & 0.03 & \nodata \\
T26~Ba,Bb & $ 0.066\pm0.005$ & $177.9\pm1.8$ & $0.06\pm0.32$ & $0.17\pm0.16$ & 0.97 & 90 \\[0.8ex]

T31~A,Ba & $ 0.652\pm0.001$ & $178.7\pm0.1$ & $1.74\pm0.01$ & $1.22\pm0.01$ & 0.16 & \nodata \\
T31~Ba,Bb & $ 0.100\pm0.001$ & $231.0\pm0.3$ & $0.61\pm0.02$ & $0.65\pm0.07$ & 0.7 & 170 \\[0.8ex]

T39~Aa,Ab & $ 1.242\pm0.001$ & $ 19.2\pm0.1$ & $0.31\pm0.01$ & $0.31\pm0.01$ & 0.67 & 4000 \\
T39~Aa,B & $ 4.497\pm0.001$ & $ 70.8\pm0.1$ & $0.77\pm0.02$ & $0.87\pm0.02$ & 0.34 & \nodata \\
\\[-2ex]
\multicolumn{7}{c}{\dotfill Unrelated background stars \dotfill} \\
T54~A,c\tablenotemark{d}     & $ 2.383\pm0.007$ & $ 81.10\pm0.15$ & $7.06\pm0.15$ & $7.13\pm0.15$ & \nodata & \nodata \\
CHX~18N~A,b\tablenotemark{d} & $ 2.738\pm0.006$ & $356.3\pm0.1$ & $5.40\pm0.03$ & $5.25\pm0.03$ & \nodata & \nodata
\enddata
\tablenotetext{a}{Random uncertainties are listed. An additional systematic uncertainty of 0.226\% is present.}
\tablenotetext{b}{Random uncertainties are listed. An additional systematic uncertainty of 0.5\degr\ is present.}
\tablenotetext{c}{Estimate based on $\Delta K_{\rm s}$ using the models of \citet{dantona97}, see \S\ref{sect:masses} for more detail.}
\tablenotetext{d}{$\Delta H$ and $\Delta K_{\rm s}$ shown were measured in 2007 follow-up imaging data. Contrasts of $\Delta J=1.65\pm0.06$, $\Delta J=7.6\pm0.3$, and $\Delta J=5.28\pm0.07$ were further measured for T54~B, T54~c, and CHXR~18N~b, respectively.}
\end{deluxetable}

\begin{deluxetable}{lcccccccccc}
\tablewidth{0pt}
\tabletypesize{\small}
\tablecolumns{11}
\tablecaption{Properties of triple systems \label{tbl:tri}}
\tablehead{
\colhead{} & \colhead{} & \multicolumn{3}{c}{\dotfill Tight subsystem \dotfill} & \colhead{} & \multicolumn{3}{c}{\dotfill Wide subsystem \dotfill} & \colhead{} & \colhead{} \\
\cline{3-5}\cline{7-9}
\colhead{Name} & \colhead{Configuration} & \colhead{$\theta_t (\arcsec)$} & \colhead{$q_t$} & \colhead{$P_t^\prime$ (yr)} & \colhead{} & \colhead{$\theta_w (\arcsec)$} & \colhead{$q_w$} & \colhead{$P_w^\prime (yr)$} & \colhead{} & \colhead{$P_w^\prime/P_t^\prime$}
}
\startdata
CHXR 28  & (Aa,Ab),B & 0.143 & 0.73 &  100 &  & 1.85 & 0.45 &  4000 &  &  40 \\
CHXR 68A & (Aa,Ab),B & 0.101 & 0.84 &   60 &  & 4.41 & 0.32 & 16000 &  & 260 \\
CHXR 9C  & A,(Ba,Bb) & 0.130 & 0.46 &  170 &  & 0.89 & 0.92 &  2000 &  &  12 \\
T26      & A,(Ba,Bb) & 0.066 & 0.97 &   90 &  & 4.59 & 0.05 & 12000 &  & 130 \\
T31      & A,(Ba,Bb) & 0.100 & 0.70 &  170 &  & 0.68 & 0.27 &  1400 &  & 8.2 \\
T39      & (Aa,Ab),B & 1.242 & 0.67 & 4000 &  & 4.21 & 0.20 & 21000 &  & 5.3
\enddata
\end{deluxetable}

\clearpage

\twocolumngrid

\noindent possible configurations, (Aa,Ab),B and A,(Ba,Bb), are observed three times each. 

A pseudo orbital period $P^\prime$ was calculated for each system assuming that the actual orbital semi-major axes are equal to the projected separations observed. The mass ratios of the companions and the pseudo orbital periods are indicated in Tables~\ref{tbl:mult} and \ref{tbl:tri}. Three binary systems have a pseudo-orbital period of $\sim$50~yr and thus for these systems, depending on their orbital phase and orientation, a partial orbit could be measured within the next 10--20 years to derive dynamical constraints on their masses. The tight pair of CHXR~68A has the shortest pseudo-period among the triple systems observed and could allow a partial orbit determination within 10--20 years.

\subsection{Probability of chance alignment}\label{sect:prob}

We have used star counts based on the Two Micron All-Sky Survey (2MASS) Point Source Catalog (PSC) to estimate the probabilities that the candidate companions identified in our images are unrelated background stars rather than real companions; a similar approach was used by \citet{correia06}. For each target in our sample, we have retrieved all sources from the 2MASS PSC that lie within a radius of 15\arcmin\ and we have built a cumulative distribution of the number of sources as a function of $K_{\rm s}$ magnitude. These cumulative distributions were then converted to surface densities. For most of our targets these surface densities are slight overestimates as their immediate vicinity is usually less densely populated, owing to higher extinction, than their broader surroundings extending to 15\arcmin. Based on these surface densities, we have calculated, for each of the 44 candidate companions identified, the probability that a background star at least as bright as the companion and within the same angular separation would have been found, and detected, around each of the 126 targets. For the vast majority of candidates and targets, these probabilities are below $10^{-4}$. Then for each candidate companion, by summing these background source detection probabilities over the 126 target stars, we obtained the expected total number of background sources of the same brightness 
or brighter and within the same angular separation that should have been detected by our survey.

For all but 5 of the candidate companions, this expected number is at the level of 0.03 or lower, indicating that the candidate companions are almost certainly bound to their primary. The 5 candidate companions with higher expected numbers are the faint tertiary around T54 (expected number of equivalent background sources detected of 0.5), the faint source around CHXR~18N (0.4), the source around T6 (0.25), the tight binary system around T26 (0.25), and the most distant source around T39 (0.08). Given these expected numbers, denoted $\lambda$, and assuming Poisson statistics, the probabilities that the distant companions are truly bound objects are simply given by $e^{-\lambda}$, i.e. the probability to detect zero equivalent background source. For the above candidate companions, these probabilities are 0.60, 0.67, 0.78, 0.78, and 0.92, respectively. The five sources with the lowest probabilities of being bound are discussed in more detail in the next section; all other candidates are considered bound for the present study.

The above analysis for the binary systems CHXR~68A/CHXR~68B and Hn~21W/Hn~21E gives expected numbers of equivalent background stars of 0.08 and 0.4, respectively. However, as both components of these systems are confirmed members of Cha~I and the surface density of members of Cha~I is much smaller than the surface density of foreground/background stars, the above analysis does not apply to these systems and we consider them as almost certainly physically bound.

\subsection{Follow-up of the widest companion candidates}\label{sect:followup}

In this section, additional astrometric and spectroscopic data are used to establish the nature of the five candidate companions with the lowest probabilities of being bound based on stellar surface density statistics. This analysis is based on the imaging and spectroscopic follow-up of the sources T54 and CHXR~18N that was mentioned previously, as well as on data retrieved from the {\em HST} archive and from the literature.

We have retrieved and analyzed {\em HST/WFPC2} images for the stars CHXR~18N, T6 and T26. In some of those images, the peak of the primary star PSF is saturated and thus cannot be used for centroid determination. In those cases, the centroid of the primary star was found by maximizing the cross-correlation of the secondary mirror support diffraction spikes after a 180\degr\ rotation of the image. The candidate companion around CHXR~18N is also saturated in the HST/WFPC2 images, but its diffraction spikes are too faint to be used for PSF registration; in this case, the centroid was determined by cross-correlation, within an annulus excluding the saturated pixels, with the PSF of an unsaturated nearby star in the image. In all other cases, i.e. for unsaturated PSFs, the centroid was determined by fitting a 2-D Gaussian function. For this analysis, the tight binary around T26 was treated as a single object and measurements were made relative to its center of light. While not resolved in the {\em HST} images, the PSF of this tight binary shows hints of elongation. For the candidate companion T39~B, we have used relative astrometry measurements reported in \citet{correia06}, based on VLT/NACO observations. All of the astrometric measurements are summarized in Table~\ref{tbl:followup}.

\begin{figure*}[t!]
\epsscale{0.95}
\plotone{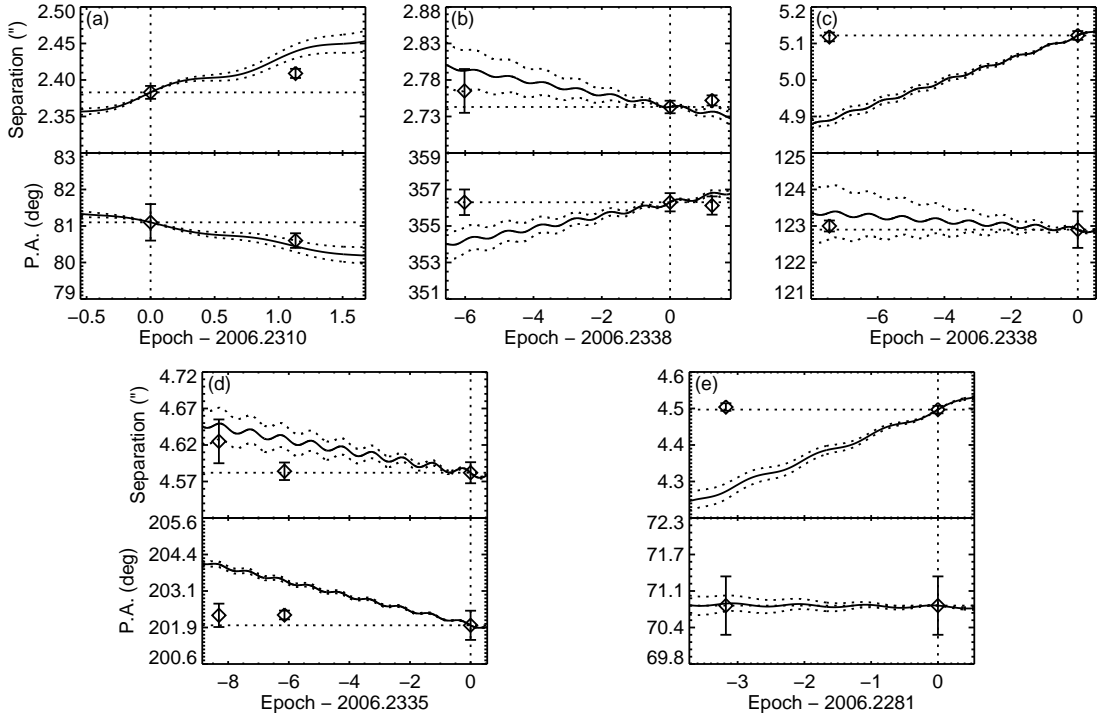}
\caption{\label{fig:pm} Relative positions of the candidate companions T54~c ({\it a}), CHXR~18N~b ({\it b}), T6~B ({\it c}), T26~BaBb ({\it d}) and T39~B ({\it e}) as a function of time. Open diamonds mark the observed separations ({\it top}) and position angles ({\it bottom}) of the candidate companions at the epochs for which we have measurements. The solid lines indicate the expected separation and position angle of a stationary background source as a function of time; the dotted lines reflect the uncertainties on the proper motion and distance of the primary stars.}
\end{figure*}

Figure~\ref{fig:pm} shows a comparison of the measured separations and position angles of the five candidate companions with those expected for stationary background stars based on the proper motions of their primary. The proper motions of the primary stars were taken from the UCAC2 catalog \citep{ucac2}, except for the star T54 for which it was taken

\onecolumngrid
\begin{deluxetable}{lcccc}
\tablewidth{0pt}
\tabletypesize{\footnotesize}
\tablecolumns{5}
\tablecaption{Multiple epoch astrometry of the widest candidate companions \label{tbl:followup}}
\tablehead{
\colhead{System} & \colhead{Epoch} & \colhead{Separation\tablenotemark{a}} & \colhead{Position Angle\tablenotemark{a}} & \colhead{Source} \\
 & \colhead{(yr)} & \colhead{(\arcsec)} & \colhead{(\degr)} & }
\startdata
T54~A,c & 2006.23098 & $2.383\pm0.007\pm0.005$ & $81.10\pm0.15\pm0.5$ & 1  \\
        & 2007.36340 & $2.409\pm0.003\pm0.005$ & $80.62\pm0.07\pm0.2$\tablenotemark{b} & 1 \\
\\[-2ex]
CHXR~18N~A,b & 2000.21481 & $2.76\pm0.03$ & $356.3\pm0.7$ & 2 \\
             & 2006.23382 & $2.738\pm0.006\pm0.006$ & $356.3\pm0.1\pm0.5$  & 1 \\
             & 2007.46197 & $2.747\pm0.003\pm0.006$ & $356.12\pm0.06\pm0.5$  & 1 \\
\\[-2ex]
T6~A,B & 1998.78593 & $5.118\pm0.012$         & $123.00\pm0.15$ & 2\\
       & 2006.23380 & $5.122\pm0.003\pm0.011$ & $122.9\pm0.1\pm0.5$ & 1 \\
\\[-2ex]
T26~A,(Ba,Bb)\tablenotemark{c} & 1997.92682 & $4.63\pm0.03$           & $202.3\pm0.4$ & 2\\
                               & 2000.09098 & $4.589\pm0.012$         & $202.30\pm0.15$ & 2\\
                               & 2006.23354 & $4.587\pm0.010\pm0.010$ & $202.0\pm0.1\pm0.5$ & 1 \\
\\[-2ex]
T39~Aa,B & 2003.05233 & $4.504\pm0.002\pm0.010$ & $70.8\pm0.1\pm0.5$ & 3 \\
         & 2006.22815 & $4.497\pm0.001\pm0.010$ & $70.8\pm0.1\pm0.5$  & 1 
\enddata
\tablenotetext{a}{When two uncertainties are quoted, the first is the measurement uncertainty and the second is the systematic uncertainty.}
\tablenotetext{b}{The field of view orientation of the 2007 images was calibrated relative to the 2006 images using the inner binary of this system; a correction of -0.6\degr\ was applied to the second epoch orientation obtained from the image header. The systematic uncertainty on the position angle of the second epoch measurements thus corresponds to the random uncertainty of the position angle of the inner binary.}
\tablenotetext{c}{Relative position for the center of light of component Ba,Bb.}
\tablerefs{(1) This work; (2) {\em HST/WFPC2} archive; (3) \citealt{correia06}.}
\end{deluxetable}
\twocolumngrid

\clearpage

\begin{figure}
\epsscale{1}
\plotone{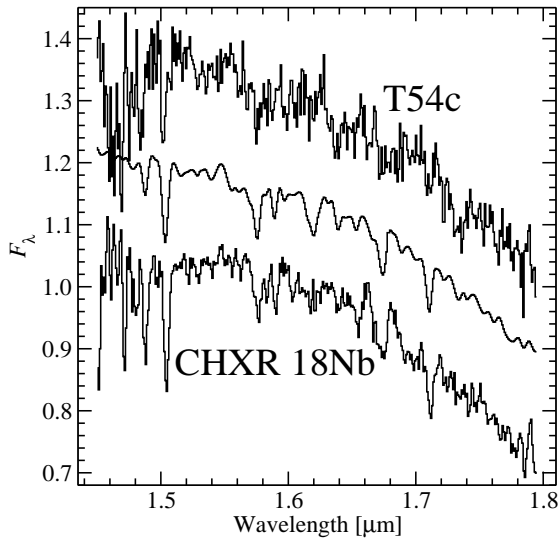}
\caption{\label{fig:spectra} $H$-band spectra of the companion candidates CHXR~18N~b and T54~c. The spectra are normalized at 1.6~$\mu$m and offset by 0.15 for clarity. The middle spectrum is an observed M1.5~V comparison spectrum (HD~36395) from the IRTF spectral library \citep{cushing05}, smoothened to the approximate resolution of the other spectra.}
\end{figure}

from \citet{frink98}.\footnote{A significantly different value is reported for T54 in the Tycho-2 catalog \citep{tycho2}, but we did not use this value as \citet{skiff07} pointed out in a note that the astrometry of this source in the Tycho-2 catalog was wrong.} Figure~\ref{fig:pm} clearly indicates that T6~B, T26~BaBb and T39~B are truly bound to their primaries, and we consider them as such for the rest of the analysis. However for the candidates T54~c and CHXR~18N~b, the comparison is rather inconclusive. For T54~c, the reduced $\chi^2$ values for the {\em bound} and {\em not bound} scenarios are 6.7 and 4.5, respectively, making it impossible to differentiate between them. For CHXR~18N, the reduced $\chi^2$ values are 1.2 and 6.4, respectively, revealing a better agreement with true physical association but with a low level of significance.

The spectra of the companion candidates around T54 and CHXR~18N are shown in Figure~\ref{fig:spectra}. If the companion candidates had been real companions, their K-band contrasts would have implied late-M spectral types and masses of $\sim$0.025~\msun\ and $\sim$0.015~\msun, respectively. Late M-type objects have a characteristic triangular-shaped $H$-band spectrum (Cushing et al.\ 2005) that is further emphasized by the lower surface gravity of young objects \citep{lodieu08}. We compared the spectra of the two companion candidates to the DUSTY models by \citet{chabrier00} and to observed near-infrared spectra from the IRTF spectral library \citep{cushing05}, and found the spectra of these two companion candidates to be clearly incompatible with spectral types later than M5. The best-fit spectrum from the IRTF library is also shown in Figure~\ref{fig:spectra}, and comes from the M1.5\,V star HD~36395. We conclude that the companion candidates T54~c and CHXR~18N~b are likely background dwarfs, located at $\sim$350~pc and $\sim$800~pc, respectively.

\section{Discussion}\label{sect:discussion}

\subsection{Overall multiplicity frequency}\label{sect:mfs}

The multiplicity fraction (MF) and the mean number of companions per primary star, or companion star fraction (CSF), are indicated in Table~\ref{tbl:mfs} for our raw and complete Cha~I samples. For our complete sample, defined as including only targets for which we have reached a contrast limit at least as good as the top curve (90\%) of figure~\ref{fig:comp} and counting only companions falling above this curve and in the separation range 0.1\arcsec--6\arcsec, the MF is $0.27_{-0.04}^{+0.05}$ and the CSF is $0.32_{-0.05}^{+0.06}$.

\onecolumngrid
\begin{deluxetable}{lccccccc}
\tablewidth{0pt}
\tabletypesize{\footnotesize}
\tablecolumns{8}
\tablecaption{Multiplicity fractions of various samples \label{tbl:mfs}}
\tablehead{
\colhead{Study} & \colhead{Region} & \colhead{S} & \colhead{B} & \colhead{T} & \colhead{MF} & \colhead{CSF} & \colhead{$f_3$}
}
\startdata
\multicolumn{8}{c}{\dotfill Raw sample \dotfill} \\
This work & Cha~I & 90 & 30 & 6 & $0.29_{-0.04}^{+0.05}$ & $0.33_{-0.05}^{+0.06}$ & $0.17_{-0.06}^{+0.08}$ \\
\multicolumn{8}{c}{\dotfill Complete sample\tablenotemark{a} \dotfill} \\
\\[-2ex]
This work & Cha~I & 80 & 25 & 5 & $0.27_{-0.04}^{+0.05}$ & $0.32_{-0.05}^{+0.06}$ & $0.17_{-0.07}^{+0.09}$ \\
\\[-2ex]
\multicolumn{8}{c}{\dotfill Restricted sample: $0.13\arcsec < \theta < 6.4\arcsec$, $\Delta K < 2.5$, and $K < 10.4$ \dotfill} \\
This work & Cha~I & 48 & 19 & 2 & $0.30_{-0.06}^{+0.06}$ & $0.33_{-0.07}^{+0.08}$ & $0.10_{-0.06}^{+0.11}$ \\
\citet{ratzka05} & $\rho$~Oph & 103 & 34 & 2 & $0.26_{-0.04}^{+0.04}$ & $0.27_{-0.04}^{+0.05}$ & $0.06_{-0.04}^{+0.07}$ \\
\citet{kohler00} & Sco-Cen & 59 & 27 & 2 & $0.33_{-0.05}^{+0.06}$ & $0.35_{-0.06}^{+0.07}$ & $0.07_{-0.04}^{+0.08}$ \\
\citet{kohler98} & Taurus & 50 & 19 & 1 & $0.29_{-0.06}^{+0.06}$ & $0.30_{-0.06}^{+0.07}$ & $0.05_{-0.04}^{+0.10}$ \\
Combined sample & \nodata & 260 & 99 & 7 & $0.29_{-0.02}^{+0.03}$ & $0.31_{-0.03}^{+0.03}$ & $0.07_{-0.02}^{+0.03}$ \\
\\[-2ex]
\multicolumn{8}{c}{\dotfill Restricted sample: $0.1<M/\msun <2.0$, $60 < \theta/{\rm AU} < 500$, and $\Delta K < 3.5$ \dotfill} \\
This work & Cha~I & 83 & 13 & 0 & $0.14_{-0.03}^{+0.04}$ & \nodata & \nodata \\
\citet{kohler06} & ONC & 105 & 5.4 & 0 & $0.05_{-0.02}^{+0.03}$ & \nodata & \nodata \\
\\[-2ex]
\multicolumn{8}{c}{\dotfill Restricted sample: $32 < \theta/{\rm AU} < 640$ and $q>0.1$ \dotfill} \\
This work & Cha~I & 58 & 18 & 0 & $0.24_{-0.05}^{+0.06}$ & \nodata & \nodata \\
\citet{duchene99} & IC~348 & 66 & 8.5 & 0 & $0.13_{-0.04}^{+0.05}$ & \nodata & \nodata
\enddata
\tablenotetext{a}{Considering only the 110 targets for which we have reached a contrast limit at least as good as the top curve (90\%) of figure~\ref{fig:comp} and counting only companions falling above this curve and in the separation range 0.1\arcsec--6\arcsec.}
\end{deluxetable}
\twocolumngrid

We first compare the MF and CSF in Cha~I and in three other dispersed star-forming regions using the results of earlier surveys. However, as previous surveys have typically targeted brighter stars, had lower angular resolution, and were less sensitive, a direct comparison is impossible and we are forced to consider a more limited subset of our observations. \citet{ratzka05} conducted a multiplicity survey of 158 members of the $\rho$~Ophiuchus dark cloud using speckle imaging on a 3.5m telescope; all of their targets have a $K$-band magnitude $\lesssim$10.4. Here we consider only their restricted sample, defined as comprising only those targets with high likelihood of membership in $\rho$~Oph and with flux ratio limits $<$0.1 at $\ge$0.13\arcsec, and only those companions in the separation range 0.13\arcsec--6.4\arcsec\ and $K$-band flux ratio $\ge$0.1. \citet{kohler00} conducted a multiplicity survey of 118 X-ray selected stars in the Scorpius-Centaurus OB association also using speckle imaging on a 3.5m telescope. Again we consider only their restricted sample, defined as above and with a similar range of target magnitudes. Finally, \citet{kohler98} obtained similar observations of the Taurus star-forming region. Although they do not provide an equivalent restricted sample in their paper, we have constructed one using their Tables~1 and 2. The statistics of these three samples are shown in Table~\ref{tbl:mfs}, along with the results for our Cha~I sample after applying the same restrictions. As it can be seen in this table, the MF values are very similar in all four regions: the probabilities that the MF in Cha I and in $\rho$~Oph, Sco-Cen, and Taurus were drawn from the same underlying binomial distribution are 0.51, 0.86, and 0.85, respectively.\footnote{Calculated using the ``exact'' hypothesis test for comparison of two binomial distributions detailed in Appendix~B2 of \citet{brandeker06}.} Thus it appears that for brighter targets (typically earlier than $\sim$M2-M3) and in the separation range 20--1000~AU, the multiplicity fractions are the same in most dispersed star-forming regions. Given that the four samples considered above are statistically indistinguishable, it is legitimate to combine them to obtain a more statistically significant sample of 257 singles, 97 doubles, and 8 triples; this yields MF=0.290$_{-0.024}^{+0.025}$ and CSF=$0.31_{-0.03}^{+0.03}$.

Previous multiplicity studies have revealed that the MF in dense star-forming clusters is significantly below that in more dispersed regions; we investigate this here by comparing our results with observations of the Orion Nebula Cluster (ONC) and IC~348. \citet{kohler06} surveyed 228 stars in the periphery of the ONC using adaptive optics, and they complemented their study with the speckle imaging results of \citet{petr98} for the cluster core. Given that these authors found only a small and statistically not very significant difference in binarity between the cluster core and its periphery, we consider only their combined sample that incorporates stars in both the core and the periphery. For stars with high likelihood of membership in the cluster and with masses in 0.1--2~\msun, for the separation range 60--500~AU, and for contrast below roughly 3.5~mag in $K$, they report an MF of $0.051\pm0.027$. Applying these restrictions to our Cha~I sample, we obtain MF=0.14$_{-0.03}^{+0.04}$. The probability that both MFs were drawn from the same distribution is only 0.04; the overabundance of binaries in Cha~I relative to the ONC, by a factor $\sim$$2.7$, is thus statistically significant. \citet{duchene99} observed 66 stars in the IC~348 cluster using adaptive optics; their target sample comprises stars of roughly the same masses as ours. For the orbital separation interval 32--640~AU and for mass ratios $>$0.1, they report 8.5 binaries after correction for incompleteness; this corresponds to MF=0.13$_{-0.04}^{+0.05}$. For equivalent restrictions in our Cha~I sample, we obtain MF=0.24$_{-0.05}^{+0.06}$. Thus the MF in Cha~I is also higher than that in IC~348, by a factor $\sim$$1.8$; the probability that both MFs were drawn from the same distribution is 0.07. Our observations thus support earlier findings that the MF is lower in dense star-forming clusters than in dispersed star-forming regions. The binary statistics of the ONC, IC~348, and Cha~I samples are summarized in Table~\ref{tbl:mfs}.

The ratio of the number of systems with at least three components to the number of systems with at least two components, denoted $f_3$, is another interesting quantity to compare. This value is indicated in Table~\ref{tbl:mfs} for our Cha~I sample as well as for the $\rho$~Oph, Sco-Cen, and Taurus samples described above. The probabilities that the values of $f_3$ in Cha~I and in $\rho$~Oph, Sco-Cen and Taurus were drawn from the same distribution are all above 0.6; the differences are thus not statistically significant. Combining the four samples, as done above for the MF, leads to a more statistically significant value of $f_3=0.08_{-0.03}^{+0.03}$. \citet{correia06} obtained AO imaging of 58 known wide binaries ($>0.5$\arcsec) in a few star-forming regions. For their combined sample of 31 targets in Tau-Aur, Cha~I, and $\rho$~Oph, they resolved 9 high-order multiples, leading to $f_{3w}=0.29_{-0.09}^{+0.10}$. When considering only wide binaries in our sample (19 wide binaries, of which 6 have a third component), the corresponding value is $f_{3w}=0.32_{-0.11}^{+0.14}$, indistinguishable from the value found by \citet{correia06}.

Comparison with the field population is slightly complicated as field multiplicity studies have typically concentrated their efforts on a narrow spectral type interval (e.g. \citet{duquennoy91} for G stars; \citet{delfosse04} for M stars) and have explored a much wider orbital separation interval than our present observations. A meaningful comparison with the properties of our entire sample requires first taking the initial mass function (IMF) into account. This was done by \citet{lada06} who reports an IMF-weighted MF of $\sim$0.33 for field stars; the same value was also obtained by \citet{reid97a} for the 8~pc sample. This number is based on statistics that should be complete over all orbital separations and for mass ratios $>$0.1. The restricted orbital separation range of our observations must then be taken into account. If we assume that the orbital separation distribution of all field multiple systems is the same as that reported for G-type stars by \citet{duquennoy91}, then $\sim$41\% of field multiple systems would have a separation within 16--1000~AU, and the corresponding field MF in that interval would be $\sim$0.14. Although this is a very rough comparison, it seems to indicate that within the 16--1000~AU interval, the MF in Cha~I (0.27) is higher than in the field ($\sim$0.14), by a factor $\sim$1.85. The probability that the MFs in Cha~I and in the field were drawn from the same distribution is below 0.04; in fact, it is likely to be even lower since the above assumption overestimates the number of M dwarf binaries in the 16--1000 AU range where studies reveal a deficit of wide companions among M dwarfs compared to G-type stars.

\subsection{Multiplicity as a function of mass}\label{sect:multvsmass}

\begin{figure}
\epsscale{1}
\plotone{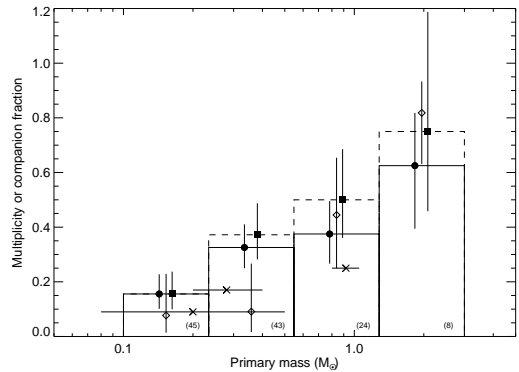}
\caption{\label{fig:multmass} Multiplicity fraction ({\it solid histogram}) and companion star fraction ({\it dashed histogram}) of our sample as function of primary mass. Open diamonds show the MF of the numerical simulations of \citet{goodwin04}, adapted to the resolution limit of our observations (see \S\ref{sect:sim} for detail). The cross symbols show field MF measurements corrected for the limited orbital separation range probed by our survey (see \S\ref{sect:multvsmass} for detail); the values are taken from \citet{delfosse04} for the leftmost point, from \citet{fischer92} and \citet{reid97a} for the middle point, and from \citet{duquennoy91} for the rightmost point.}
\end{figure}

To investigate the variation of multiplicity as a function of primary mass, we have divided the sample into 4 nearly-equal logarithmic mass bins and calculated the MF and CSF in each bin; the results are shown in Figure~\ref{fig:multmass} and summarized in Table~\ref{tbl:mfs2}. There is a clear increase in multiplicity fraction with increasing primary mass: it rises from 0.16$^{+0.07}_{-0.05}$ to 0.63$^{+0.19}_{-0.23}$ over the range $\sim$0.1--3~\msun. This trend has been observed previously both for other star-forming regions and for the field.

We first compare these multiplicity fractions with values for field stars from the M-type stars survey of \citet{delfosse04}; the visual multiplicity survey of 58 M2--M4.5 stars done by \citet{fischer92}, counting only the companions they found in the 10--1000~AU range; the multiplicity survey of 41 M2--M4.5 stars done by \citet{reid97a} using {\em HST}, sensitive to companions in the range 16--1000~AU; and the G-type stars survey of \citet{duquennoy91}. As the surveys of \citet{delfosse04} and \citet{duquennoy91} are complete over all orbital separations, the MF values they have reported must be corrected for the range of orbital separations probed by our observations ($\sim$12--1000~AU) to make a meaningful comparison. Based on the orbital period distribution of \citet{duquennoy91}, $\sim$44\% of G-type multiple systems have a semi-major axis in the 12--1000~AU interval, while the same orbital separation range includes roughly 1/3 of the M dwarf multiple systems in \citet{delfosse04} (see their Fig.~4). We have thus multiplied the MF values reported in \citet{delfosse04} and \citet{duquennoy91} by 0.33 and 0.44, respectively. The field values from these four studies are summarized in Table~\ref{tbl:mfs2} and shown in Figure~\ref{fig:multmass}, which indicates that the increase of the MF as a function of primary mass in Cha~I is similar to that of the field, although the values themselves are systematically higher in Cha~I.

A few recent high angular resolution multiplicity surveys of star-forming regions have targeted only low-mass stars, and their results are suitable for comparison with our results here. The Keck speckle interferometry survey of \citet{konopacky07} and the {\em HST} survey of \citet{kraus06} have targeted a combined sample of 15 Taurus members in the 0.1--0.2~\msun\ range and found 3 binaries within the separation range probed by our survey. The corresponding MF is reported in Table~\ref{tbl:mfs2} along with the value in Cha~I for the same mass range. Both MF values are in very good agreement; the probability that they were drawn from the same distribution is 0.67. \citet{bouy06} surveyed low mass stars in the Upper Scorpius star-forming region using AO at the VLT. We consider only their subsample of 35 stars with spectral type M0--M6, corresponding to $\sim$0.15--0.6~\msun\ at an age of 5~Myr according to the models of \citet{dantona97}. The comparison of this sample to ours (see Table~\ref{tbl:mfs2}) suggests a possible possible higher incidence of binaries among the low-mass stars in Cha~I compared to Upper Scorpius; the probability that the MFs in both regions, over this mass range, were drawn from the same distribution is 0.11.

In a recent study of multiplicity over the separation range 330--1650~AU, \citet{kraus07} have found a similar trend of increasing binary fraction with increasing primary mass in Upper Scorpius~A, Taurus and Cha~I. In particular for Cha~I, these authors determined binary fractions that increase from a few percents at 0.1--0.2~\msun\ to $\sim$30\% at 1.16-2.5~\msun. The observed dependence of binary fraction on primary mass thus extends beyond the separation range probed by our survey.

As can be seen from Figure~\ref{fig:multmass}, among our sample there are no triple systems in the lowest mass bin and there is a marginal increase in the fraction of triple systems with primary mass, although this trend is hampered by low number statistics in the highest mass bin. Such an increase is also observed by \citet{correia06} for their AO survey of wide binaries in various star-forming regions.

\onecolumngrid
\begin{deluxetable}{lccccccc}
\tablewidth{0pt}
\tablecolumns{7}
\tabletypesize{\scriptsize}
\tablecaption{Multiplicity fractions for different regions and different masses \label{tbl:mfs2}}
\tablehead{
\colhead{Study} & \colhead{Region} & \colhead{Mass (\msun)} & \colhead{$S$} & \colhead{$B+T+\dots$} & \colhead{MF} & \colhead{CSF}
}
\startdata
This work & Cha~I & 0.10--0.23 & 38 &  7+0 & 0.16$_{-0.05}^{+0.07}$ & 0.16$_{-0.06}^{+0.08}$ \\
This work & Cha~I & 0.23--0.55 & 29 & 12+2 & 0.33$_{-0.08}^{+0.08}$ & 0.37$_{-0.09}^{+0.12}$ \\
This work & Cha~I & 0.55--1.28 & 15 &  6+3 & 0.38$_{-0.11}^{+0.12}$ & 0.50$_{-0.14}^{+0.19}$ \\
This work & Cha~I & 1.28--3.00 &  3 &  4+1 & 0.63$_{-0.23}^{+0.19}$ & 0.75$_{-0.29}^{+0.44}$ \\
\\[-2ex]
\citet{delfosse04}& Field & 0.08--0.5& \nodata   & \nodata & $\sim$0.09 & \nodata \\
\citet{fischer92} & Field & 0.2--0.4 &  48 &  10 & $0.17_{-0.05}^{+0.06}$ & \nodata \\
\citet{reid97a}   & Field & 0.2--0.4 &  34 &   7 & $0.17_{-0.06}^{+0.08}$ & \nodata \\
\citet{duquennoy91}& Field& 0.8--1.05& \nodata & \nodata & $\sim$0.25 & \nodata \\
\\[-2ex]
\citet{konopacky07}& Taurus & 0.1--0.2 & 12 & 3 & $0.20_{-0.11}^{+0.15}$ & \nodata \\
and \citet{kraus06}& & & & & & \\
This work         & Cha~I  & 0.1--0.2 & 33 & 5 & $0.13_{-0.05}^{+0.08}$ & \nodata \\
\\[-2ex]
\citet{bouy06}    & Up Sco & 0.15--0.6 & 30 & 5 & $0.14_{-0.06}^{+0.08}$ & \nodata \\
This work         & Cha~I & 0.15--0.6 & 65 & 26 & $0.29_{-0.05}^{+0.05}$ & \nodata \\
\enddata
\end{deluxetable}
\twocolumngrid

\subsection{Mass ratio as a function of primary mass}

\begin{figure}
\epsscale{1}
\plotone{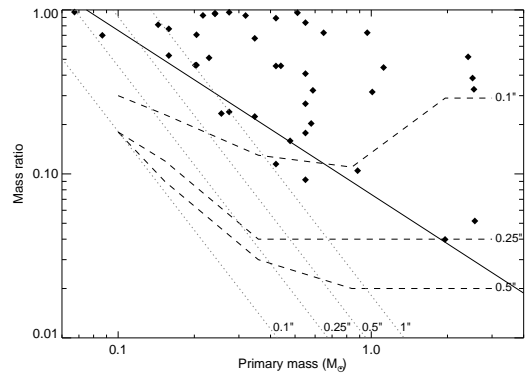}
\caption{\label{fig:qvsm} Mass ratios of the binary systems identified in our survey as a function of the primary mass. The solid line shows the mass ratio for a companion of 0.075~\msun. The dashed lines indicate the detection sensitivities reached for at least 50\% of the targets for different angular separations. The dotted lines indicate a constant binding energy of $2\times 10^{42}$~erg for various orbital separations, see \S\ref{sect:sepvsm} for more detail. Typical uncertainties on the estimated masses and mass ratios are $\sim$15\% and $\sim$25\%, respectively.}
\end{figure}

Figure~\ref{fig:qvsm} shows the mass ratio as a function of primary mass for all the multiple systems found. For this diagram, and the similar ones to come, each triple system contributes two points: one for the tight subsystem and one for the wide system. The mass ratios and separations for these are indicated in Table~\ref{tbl:tri}. Figure~\ref{fig:qvsm} clearly shows that as the primary mass decreases, the multiple systems are restricted to increasingly higher mass ratios. While such a trend could result from a selection effect, as lower mass ratio companions are more difficult to detect around lower mass primaries, this is not the case here as our detection limits would have allowed us to detect systems with mass ratios much below the observed lower envelope at the low-mass end. The same tendency is also apparent among field multiple stars; this was recognized rather early based on a comparison of the mass ratio distributions of G-type stars, early-M stars, and late-M stars and brown dwarfs \citep[e.g.][]{duquennoy91,fischer92,close03}.

Interestingly, the lower envelope of the mass ratio distribution corresponds to a constant mass, which in fact roughly coincides with the stellar/substellar boundary of $\sim$0.075~\msun. In Cha~I, according to our mass determinations, there are only a handful of companions lying just below the substellar boundary. However, given the uncertainties in mass and mass ratio determinations (see \S\ref{sect:masses}), it is possible that most or all of these companions have stellar masses. This low incidence of brown dwarf companions to stars is also found for G--M field stars, for which high-contrast imaging surveys revealed an incidence of brown dwarf companions of $\sim$2--7\% for the separation range 25--1600~AU \citep{metchev06, lafreniere07}. This suggests that the process through which companions form at separations greater than $\sim$15~AU around stars more massive than $\sim$0.1~\msun\ is inefficient in the substellar mass regime. However, the situation may be different for very low-mass stars ($<0.1$~\msun), for which several systems harboring a brown dwarf companion are known.\footnote{See the Very Low Mass Binaries Archive at http://paperclip.as.arizona.edu/~nsiegler/VLM\_binaries/}

Figure~\ref{fig:qvsm} also reveals a dearth of near-equal mass binaries at the high mass end: the six systems with primary mass $>1$~\msun\ have a mass ratio below 0.55. This is not the result of an observational bias as high mass ratios would be easier to detect. The situation is similar for the high-mass binary systems in Scorpius~OB2, which also show a preference for low mass ratios \citep{kouwenhoven05}. The same trend is also found in the study of field G-type binary systems \citet{duquennoy91}. The findings of \citet{kraus07}, for their combined Cha~I, Taurus and Upper Scorpius sample, do not display this trend (see their figure~7); their sample contains $\sim$12 binary systems with primary masses 1.16--2.5\msun\ and $q>0.75$. However, of these $\sim$12 systems, more than half come from Upper Scorpius while only two come from Cha~I. Moreover, one of these two Cha~I systems (T26) clearly has a mass ratio much below 0.75: the 2MASS-based marginally resolved photometry used by \citet{kraus07} to derive its mass ratio differs by $\sim$3~mag from ours and from values reported elsewhere in the literature. Taking these factors into consideration, their results are in agreement with ours, but may point toward a difference in high-mass binary properties between Cha~I and Upper Scorpius.

\subsection{Mass ratio as a function of orbital separation}

\begin{figure}
\epsscale{1}
\plotone{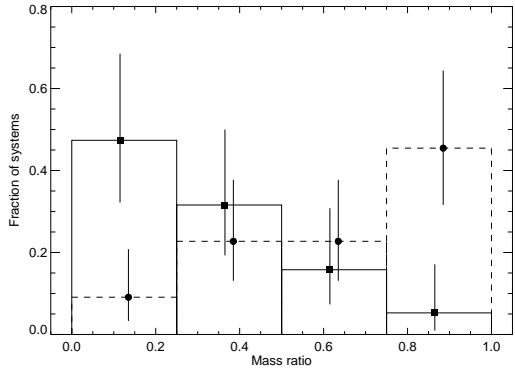}
\caption{\label{fig:qdist-sep} Mass ratio distributions for binaries with separation above ({\it solid histogram}) and below ({\it dashed histogram}) 50~AU. There is roughly an equal number of objects in both samples.}
\end{figure}

Among our Cha~I sample, two main features can be noted regarding the distribution of mass ratio versus separation. First, there is a clear lack of near-equal mass binaries ($q>0.5$) at large separations ($>50$~AU) even though such systems would be very easy to detect. Second, closer binaries tend to have mass ratios closer to unity. Both of these trends can be seen on Figure~\ref{fig:qdist-sep}, which shows the mass ratio distributions for the subsets of small ($<50$~AU) and large ($>50$~AU) orbital separations. The first mass ratio bin of the small separation sample is the only one that could be significantly affected by incompleteness (see Table~\ref{tbl:qlim}). By considering only systems with a mass ratio over 0.2, where incompleteness should be unimportant, a Kolmogorov-Smirnov test indicates that the probabilities that the mass ratios of the large- and small-separations samples were drawn from a uniform distribution are 0.03 and 0.14, respectively, lending good statistical significance to the first trend mentioned above but only marginal significance to the second. The probability that the mass ratios for the small- and large-separation samples were drawn from the same distribution is only 0.02, again considering only systems with a mass ratio above 0.2.

\citet{ratzka05} looked at the flux ratio distributions for close ($<180$~AU) and wide ($>180$~AU) companions in $\rho$~Oph (see their Fig.~13). As the flux ratio is a good proxy of mass ratio, their findings can be compared to ours. Before proceeding further, we note that both features mentioned above are still be apparent if the separation between close and large separation systems is made at 180~AU instead of 50~AU, although the statistical significance is much reduced as there are only 12 systems with a separation greater than 180~AU. For the wide companions \citet{ratzka05} observed the same trend as us, namely, that wide pairs preferentially have large flux differences. However, the flux ratio distribution for their close pairs is more or less uniform, and does not show a peak near unity as for our sample. Similarly in Sco-Cen, \citet{kohler00} found a slight preference for wide pairs ($>190$~AU) to have large flux differences, while the flux ratio distribution of close pairs is rather uniform. In Taurus, the wide pairs ($>180$~AU) display a preference for large flux differences, and the close pairs also show a peak for flux ratios near unity \citep{kohler98}, very similar to our findings in Cha~I.

\subsection{Orbital separation as a function of mass}\label{sect:sepvsm}

\begin{figure}
\epsscale{1}
\plotone{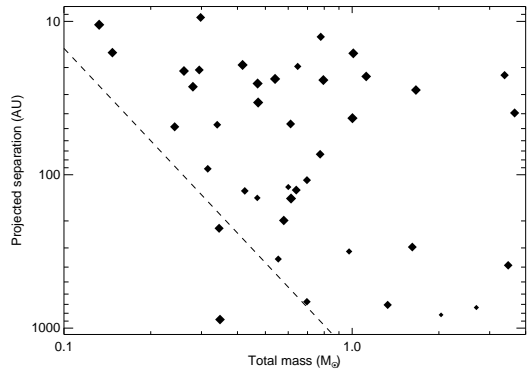}
\caption{\label{fig:sepvsm} Projected separations of the binary systems identified in our survey as a function of their total masses. The sizes of the symbols are proportional to the logarithm of the mass ratios of the binaries. The dashed line corresponds to a separation equal to $1500\,(M_{\rm tot}/\msun)^{2}$~AU. Typical uncertainties on the estimated total masses are $\sim$20\%.}
\end{figure}

Figure~\ref{fig:sepvsm} shows the projected separations of the binary systems as a function of their total masses. It clearly shows a lack of wide systems at low masses; this is not an observational bias as companions with larger separations are easier to detect. The observed decline of the maximum orbital separation of Cha~I binaries with decreasing mass is consistent with earlier findings for field systems \citep[e.g.][]{burgasser03, close03}. The envelope of the separation distribution of our Cha~I sample is approximated reasonably well by $1500\,(M_{\rm tot}/\msun)^2$~AU; only one binary, Hn~21W at 0.34~\msun\ and 880~AU, departs significantly from this trend. A very similar relation [$1400\,(M_{\rm tot}/\msun)^2$~AU] was previously suggested for low-mass binary systems in the field \citep{burgasser03}, with an apparent break at $\sim$0.6~\msun, above which wider pairs are found. For Cha~I it is not possible, based on our data, to assess whether this relation is valid for total masses greater than $\sim$0.8~\msun\ since the corresponding orbital separations are too close to or above our upper limit of $\sim$1000~AU. As the above relation between orbital separation and mass squared is reminiscent of a constant binding energy cutoff, we have calculated the binding energies of all binary systems. As suggested by the above trend, all systems (except Hn~21W) have binding energies above $\sim$$2\times 10^{42}$~erg. As before, this boundary is meaningful only for total masses below $\sim$0.8~\msun\ as our observations were much less sensitive to binaries with lower binding energies beyond this mass. Also, the uncertainty on this value is quite large, up to 50--60\%, due to the uncertainties in masses, mass ratios, and projection effects.

The apparent lower boundary of binding energy could give rise to a trend such as the one noted earlier that less massive binaries preferentially have mass ratios closer to unity. Indeed, for low primary masses, the occurrence of companions of even lower masses could be suppressed as these would have binding energies that were too low. To investigate this possibility we have over plotted lines of constant binding energy ($\sim$$2\times 10^{42}$~erg) on figure~\ref{fig:qvsm} for four angular separations. Comparison of these lines with the data indicates that for separations below 1\arcsec, 0.5\arcsec, and 0.25\arcsec, and primary masses above 0.75~\msun, 0.3~\msun, and 0.2~\msun, respectively, systems with mass ratios much lower than the minima observed would still have binding energies above the minimum observed. Thus it does not appear that the envelope of the mass ratio--primary mass distribution is dictated by binding energy. Rather, this diagram seems to be better explained by a fixed minimum mass for companions. The binding energy appears to be responsible only for the decrease of the orbital separation with decreasing mass.

\subsection{Stability of triple systems}

\citet{tokovinin04}, by studying a sample of triple systems from the Multiple Star Catalog, determined the following empirical stability criterion, $P_{\rm w}/P_{\rm t} (1-e_{\rm w})^3 > 5$, and an upper cutoff on the periods ratios of $P_{\rm w}/P_{\rm t}<10^4$, where $P_{\rm w}$ and $P_{\rm t}$ are the periods of the wide and tight subsystems and $e_{\rm w}$ is the eccentricity of the wide subsystem. This criterion cannot be verified for our triple systems as their true orbital separations and eccentricities are unknown. Nevertheless, the pseudo-periods calculated above (see Table~\ref{tbl:tri}) can provide some useful information. At first glance, all systems appear stable as they all have a pseudo-period ratio that satisfies the empirical stability criterion. Three systems (CHXR~28, CHXR~68A, and T26) have pseudo-period ratios well above the minimum value of 5, indicating that they are almost certainly stable hierarchical triple systems. The other three systems, CHXR~9C, T31, and T39, have ratios of 12, 8.2, and 5.3, respectively, and thus could be unstable depending on their orbital orientations and eccentricities. 

We have performed a simple Monte Carlo simulation to investigate the likelihood of stability of these three triple systems. First we have randomly defined $10^5$ fiducial stable triple systems filling uniformly the stability region of the $e_{\rm w}$ vs $\log\left[P_{\rm w}/P_{\rm t} (1-e_{\rm w})^3\right]$ plane. The tight and wide pairs of each fiducial system were then randomly assigned an orbital orientation, phase, and eccentricity, and the corresponding projection factors were calculated. The eccentricities of the wide pairs were drawn from a distribution following $f(e) \propto 2e$. Then using the projection factors, a pseudo-period ratio was derived for each fiducial stable system. Of all fiducial stable triple systems, only 6.3\%, 4.1\%, and 3\% have a pseudo-period ratio smaller than 12, 8.2, and 5.3, respectively. Based on these fractions, it appears possible that one or more of CHXR~9C, T31, and T39 is indeed unstable; however we cannot reach a definite conclusion.

\subsection{Comparison with numerical simulations}\label{sect:sim}

In this section we compare our observations mainly with the results of \citet{goodwin04} and \citet{delgado04}, which have done numerical simulations of the collapse and fragmentation of an ensemble of low-mass ($\sim$5~\msun), dense, turbulent prestellar cores. \citet{goodwin04} considered a low level of turbulence, and explored the effect of higher levels of turbulence in a subsequent paper \citep{goodwin04b}; they have followed the evolution of the system for 0.3~Myr. \citet{delgado04} considered a level of turbulence about 20 times higher than \citet{goodwin04}, and after a hydrodynamical phase lasting $\sim$0.5~Myr, they have followed the dynamical evolution of the systems for another 10~Myr using N-body simulations. Other differences between both sets of simulations include the initial Jeans number, the initial density profile, and the equation of state. We caution the reader that, while these simulations are state-of-the-art and of very good quality, the limited range of core parameters they cover and the limited statistics they provide may impact the validity of our comparison to the entire population of Cha~I, but some interesting observations can be drawn nonetheless.

The simulations of \citet{goodwin04} yield an overall MF of 0.26 and a CSF of 0.47. Apart from an unstable quintuple, they do not form systems of multiplicity higher than 4. In fact, the two quadruple systems they form would appear to us as triple or binary given our resolution limit. \citet{goodwin04} provide the separations and masses of all of their multiple systems, it is thus possible to interpret their results within the regime where our observations were sensitive. By treating systems with separations $<$16~AU as single and ignoring single stars less massive than 0.1~\msun, we obtain a sample of 31 singles, 15 doubles, and 1 triple; this gives an MF of 0.34 and a CSF of 0.36. These do not differ significantly from the quantities mentioned above for Cha~I (0.27 and 0.32 for the same separation limits). Among the high-order multiple systems in their sample, all but one of the 10 binary subsystems have a separation less than 16~AU and half have a separation less than 10~AU. It will be interesting to see, with future radial velocity surveys or higher angular resolution observations, whether many of the components of the multiple systems in Cha~I are indeed unresolved binaries as in the \citet{goodwin04} simulations. A higher level of turbulence leads to the formation of a larger number of stars per core \citep{goodwin04b} and as a result, the average number of components per multiple system is higher, although the MF is not affected much as a larger number of single stars are also ejected from the cores. The results of \citet{goodwin04b} indicate that the raw CSF may rise from 0.47 to 0.76 when turbulence is increased by a factor of 5 compared to their baseline simulation of \citet{goodwin04}. However, applying the same procedure as above yields an MF of $\sim$0.40 and a CSF of $\sim$0.42, which is still higher than the lower turbulence case but only by a small margin. Higher levels of turbulence, such as in the simulations of \citet{delgado04}, may have an effect that can be readily compared with our results. In these simulations, even after 10.5~Myr, about 1/3 of the multiple systems are quintuples or sextuples, which seems inconsistent with our observations even if a significant number of binary subsystems were unresolved given our resolution limit. Thus this excess of high-order multiples might be regarded as an indication that the level of turbulence in Cha~I is well below that assumed by \citet{delgado04} in their simulations.

An increase of the MF with primary mass, as observed for our Cha~I sample (\S\ref{sect:multvsmass}), is also found in the numerical simulations of both \citet{goodwin04} and \citet{delgado04}; this trend was also recognized in the dynamical simulations of small clusters carried out by \citet{sterzik03}. Furthermore, the increase of high-order multiplicity with increasing primary mass we have noted earlier for our Cha~I sample is also present in both models. To relate the simulations to our observations, we have divided the systems formed in the simulations of \citet{goodwin04} into the same mass bins as for our sample and the MF was calculated within each bin by treating binaries with separation below 12~AU as single objects; the results are shown in Figure~\ref{fig:multmass}. The agreement with the observations is good, although two slight differences must be noted. First, the increase of the MF with mass is slightly steeper for the simulations than for the observations. Second, the MF in the lower mass bins are too low in the simulations. The overall behavior in the simulations of \citet{delgado04} is similar (see their Fig.~3), although the discrepancies just mentioned are more pronounced. As remarked by \citet{delgado04}, the results of these simulations are dependent on the total mass of the cloud modeled. For instance, the simulation of a more massive core would lower the MF of 2--3~\msun\ stars, whereas a lower core mass could possibly raise the MF at the low-mass end. So perhaps a better agreement with observations could be reached by simulating an ensemble of cores of different masses; this approach remains to be investigated.

The simulations of \citet{goodwin04} seem to reproduce the trend we have observed in Cha~I (Fig.~\ref{fig:qdist-sep}) that large separation binaries preferentially have smaller mass ratios: all of their systems with separation larger than 30~AU have $q<0.7$ and all of their systems with separation below 30~AU have $q>0.4$. In a similar vein, the simulations of \citet{bate00} also predict that closer binaries should have more equal mass ratios, as observed in Cha~I, owing to increased accretion on companions that form at smaller separations.

\section{Concluding Remarks}\label{sect:conclusion}

We find that the young binary population in Chamaeleon~I shares many trends with their field counterparts: the multiplicity fraction declines and the mass ratios approach unity with decreasing primary mass, the mean and maximum separation of binary systems decrease with total mass, and brown dwarfs as companions to stars are rare. Of all the properties we were able to investigate, the only clear difference between the Cha~I region and the field is the multiplicity fraction, which is lower in the field by a factor of at least $\sim$1.85 for the 16--1000~AU orbital range. Previous multiplicity surveys indicate that many properties of multiple systems, including their overabundance compared to the field, are similar in Cha~I and in other dispersed associations, but more sensitive studies are needed to investigate possible differences among them. The findings to date seem consistent with the assertion that only a small fraction of the stars in the field comes from dispersed star-forming regions, because if this were not the case a drop in multiplicity fraction with time without any effect on other properties of the multiple systems would be difficult to explain. However, these remarks may be somewhat premature given the relatively small number of multiples in our sample (36) and the lack of constraints on the very important orbital separation range below $\sim$16~AU. In any case, it will be particularly interesting to compare the same properties for multiple systems in high-density star-forming clusters, but this is a more difficult task given the larger distances to those clusters. 

Comparisons between multiplicity surveys and numerical simulations may also provide valuable insight into the star formation process. In particular, the fraction of high-order multiples could be related to the level of turbulence in the pre-stellar cores. The apparent paucity of 
such systems in our Cha~I sample could be interpreted as evidence for a low level of turbulence in this region's cores. It may be that some of the binaries and triples in our sample harbor closer-in unresolved companions. Radial velocity surveys are needed to test that possibility. 

\acknowledgments

We would like to acknowledge the excellent work of our anonymous referee, whose thorough review and thoughtful suggestions greatly helped us in improving the quality of this paper. We would like to thank our colleagues Vincent Geers, Aleks Scholz, Chris Matzner, Yanqin Wu, Andrew Shannon, and Duy Cuong Nguyen for many suggestions and enlightening discussions relating to the work presented in this paper. We also wish to thank our ESO support astronomer, Nancy Ageorges, for all the help she provided to us during the observations. DL is supported in part through a postdoctoral fellowship from the Fonds Qu\'eb\'ecois de la Recherche sur la Nature et les Technologies. This work was supported in part through grants to RJ and MHvK from the Natural Sciences and Engineering Research Council, Canada, and an Early Researcher Award to RJ from the province of Ontario.

\appendix
\section{Effect of different evolution models}\label{sect:models}

While it is not the goal of this work to compare the validity or the accuracy of the different evolution models available, it is worthwhile to investigate how a particular choice of model would influence the general trends found in our study. To address this issue, we have recalculated the masses of all primaries and companions using each of four different pre-main-sequence evolution models: \citet{dantona97}, \citet{siess00}, \citet{palla99}, and \citet{baraffe98}. When doing the analysis with each model, only the subset of objects falling within the effective temperature/magnitude range covered by that model was considered. The primary and companion masses were calculated in the way described in \S\ref{sect:masses}. For the models of \citet{palla99}, magnitudes were computed from luminosities and effective temperatures using the method described in Appendix A of \citet{testi98}; the tabulated magnitudes of the models were provided to us by F. Palla.

\vspace{0.5cm}
\twocolumngrid

\begin{figure}
\epsscale{1}
\plotone{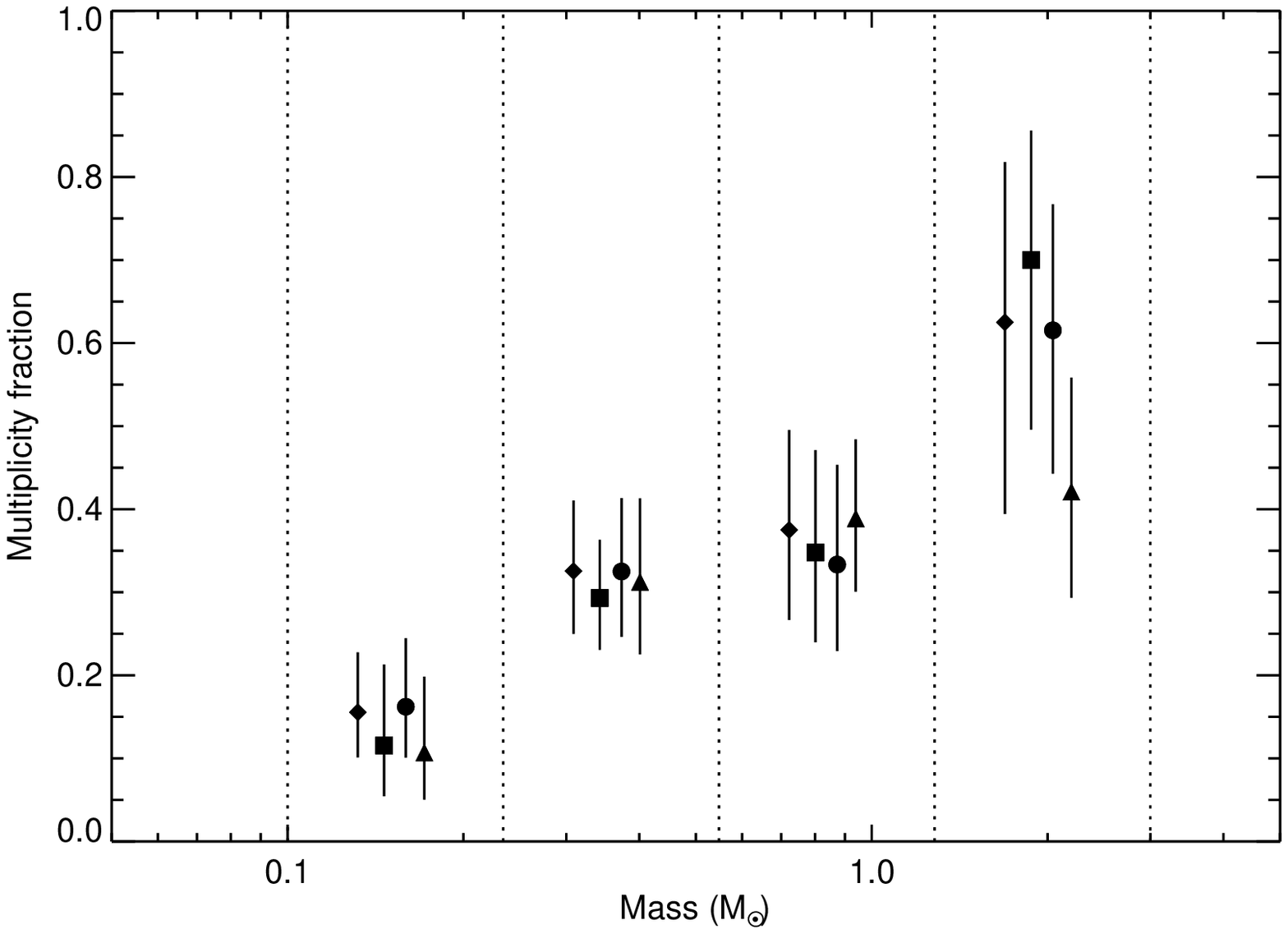}
\caption{\label{fig:multmodels} Multiplicity fraction as a function of primary mass, as calculated using different pre-main-sequence models. The dotted vertical lines mark the mass bins used. The masses and mass ratios were calculated using the models of \citet{dantona97} ({\it diamonds}), \citet{siess00} ({\it squares}), \citet{palla99} ({\it circles}), and \citet{baraffe98} ({\it triangles}); see \S\ref{sect:models} for more detail.}
\end{figure}

\twocolumngrid

\begin{figure}
\epsscale{1}
\plotone{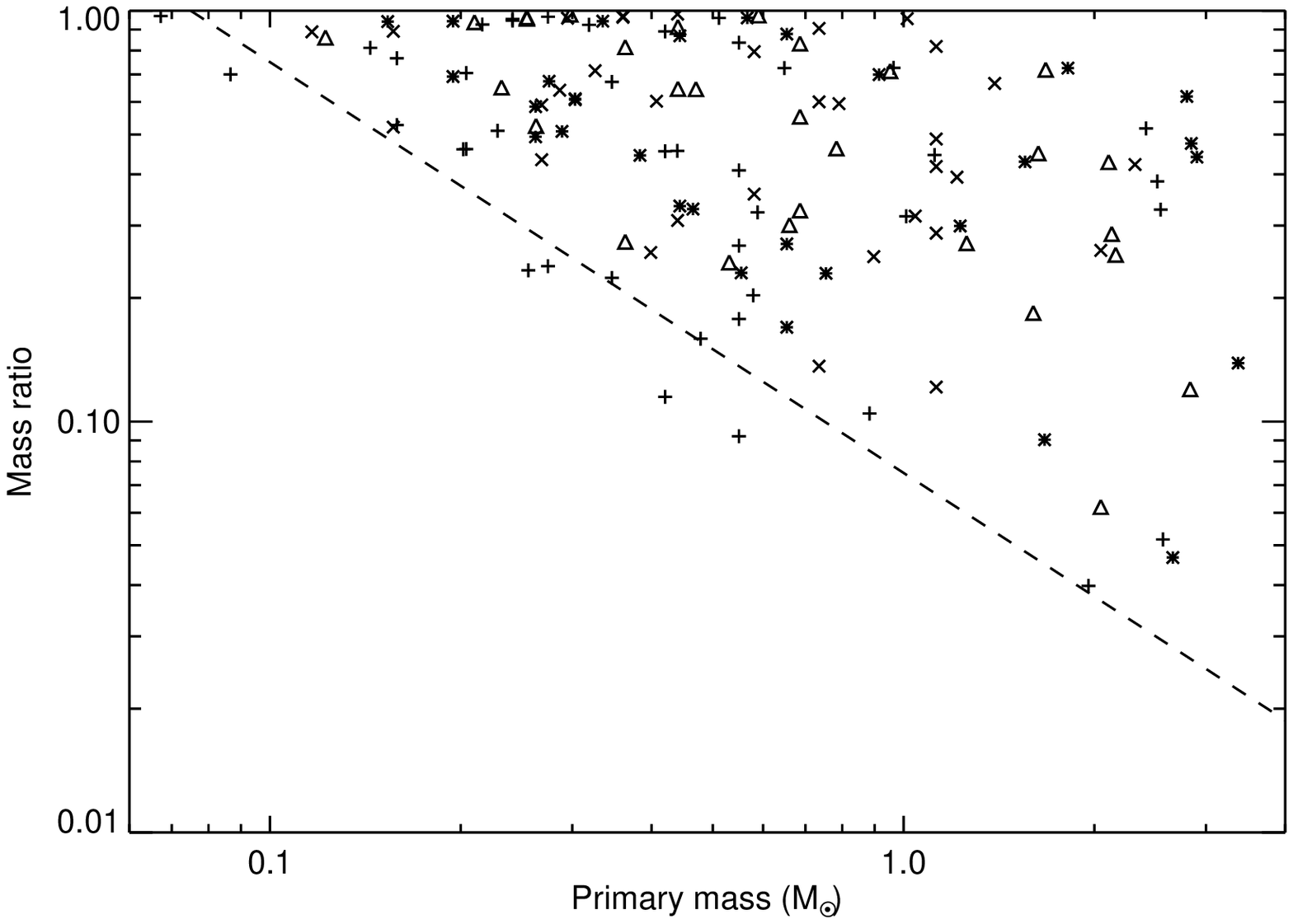}
\caption{\label{fig:qmodels} Mass ratios as a function of projected separation, as calculated using different pre-main-sequence models. The masses and mass ratios were calculated using the models of \citet{dantona97} ({\it plusses}), \citet{siess00} ({\it stars}), \citet{palla99} ({\it triangles}), and \citet{baraffe98} ({\it crosses}); see text for more detail.}
\end{figure}

\clearpage
\onecolumngrid

We concentrate on two diagnostic graphs: the MF as a function of mass, shown in Figure~\ref{fig:multmodels}, and the binary mass ratio as a function of primary mass, shown in Figure~\ref{fig:qmodels}. In computing the MF in the highest mass bin using the models of \citet{baraffe98}, since these models do not cover the whole bin, we have assumed an upper cutoff of 6000~K. Figure~\ref{fig:multmodels} shows that a very similar increase of the MF with mass is recovered using all models. The most important discrepancy occurs in the highest mass bin, where the point obtained using the models of \citet{baraffe98} is significantly below the other values. This lower MF value results from the higher (1.5--2$\times$) mass estimates obtained with the models of \citet{baraffe98} for effective temperatures above $\sim$3500~K; this shifts many single targets into the highest mass bin. Figure~\ref{fig:qmodels} indicates that a very similar distribution of mass ratio as a function of separation is also obtained using all models even though the primary masses can vary by as much as a factor of 2. Of the four sets of models considered, only those of \citet{dantona97} and \citet{baraffe98} cover the brown dwarf mass regime and a slight difference between them is worth mentioning. While the models \citet{dantona97} yield a handful of companions with masses slightly below the substellar boundary, the models of \citet{baraffe98}, which typically yield higher primary masses, do not indicate any such companions. The lower envelope of the mass ratio distribution using the models of \citet{baraffe98} corresponds to a fixed mass of $\sim$0.09~\msun\ rather than the 0.075~\msun\ value mentioned above for the models of \citet{dantona97}. Despite the small differences noted above, our analysis indicates that the specific choice of evolution model has little effect on the general multiplicity properties of the sample.


\end{document}